\documentclass[12pt,a4paper]{article}
\usepackage{graphicx}
\usepackage{cite,./mcite}
\usepackage{xspace}
\usepackage{ifthen}
\usepackage{hyperref}
\usepackage[latin1]{inputenc}
\usepackage[dvips]{color}
  \newcommand{\ccaption}[2]{
      \caption[#1]{\small{{#2}}}
  }
\begin{document}

\def\sepsyst{1}
\newcommand{\todo}[1]{\vspace*{3mm}\par\noindent\textbf{\boldmath $
\Longrightarrow$ #1}\vspace*{7mm}}
\newcommand{\madgraph}{M\scalebox{0.8}{AD}G\scalebox{0.8}{RAPH}\xspace}
\newcommand{\madevent}{M\scalebox{0.8}{AD}E\scalebox{0.8}{VENT}\xspace}
\newcommand{\ktclus}{KTC\scalebox{0.8}{LUS}\xspace}
\newcommand{\pythia}{P\scalebox{0.8}{YTHIA}\xspace}
\newcommand{\jetset}{J\scalebox{0.8}{ETSET}\xspace}
\newcommand{\amegic}{A\scalebox{0.8}{MEGIC++}\xspace}
\newcommand{\apacic}{A\scalebox{0.8}{PACIC++}\xspace}
\newcommand{\herwig}{H\scalebox{0.8}{ERWIG}\xspace}
\newcommand{\ariadne}{A\scalebox{0.8}{RIADNE}\xspace}
\newcommand{\sherpa}{S\scalebox{0.8}{HERPA}\xspace}
\newcommand{\alpgen}{A\scalebox{0.8}{LPGEN}\xspace}
\newcommand{\helac}{H\scalebox{0.8}{ELAC}\xspace}
\newcommand{\phegas}{P\scalebox{0.8}{HEGAS}\xspace}
\newcommand{\rc}{\color{red}}
\newcommand{\bc}{\color{blue}}
\newcommand{\mc}{\color{magenta}}
\newcommand{\gc}{\color{green}}
\def\revision#1{\textit{$ #1\!\!$}}
\def\fig#1{fig.~\ref{#1}}
\def\Fig#1{Fig.~\ref{#1}}
\def\figs#1{figs.~\ref{#1}}
\def\Figs#1{Figs.~\ref{#1}}

\def \sss {\scriptscriptstyle}
\def    \dRjj             {\mbox{$\Delta R_{\sss{jj}}$}}
\def\kperp{\mbox{$k_\perp$\xspace}}
\def\Eperp{\mbox{$E_\perp$\xspace}}
\def\pperp{\mbox{$p_\perp$\xspace}}
\def\ppw{\mbox{$p_{\perp W}$\xspace}}
\def\ppj{\mbox{$p_\perp^{{\rm jet}1}$\xspace}}
\def\etmin{\ensuremath{\et^{min}}}
\def\question#1{\footnote{\large\bf #1}}
\def\alps{\ensuremath{\alpha_s}\xspace}

\begin{titlepage}
\nopagebreak
  \renewcommand{\thefootnote}{\fnsymbol{footnote}}

{\begin{flushright}{
 \begin{minipage}{5cm}
   CERN-PH-TH/2007-066 \\
   LU-TP 07-13\\
   KA-TP-06-2007\\
   DCPT/07/62\\
   IPPP/07/31\\
   SLAC-PUB-12604
\end{minipage}}\end{flushright}}
\vfill
\begin{center} {\Large\bf Comparative study of various algorithms for
    the merging of parton showers and matrix elements in hadronic
    collisions~\footnote{Work supported in part by the Marie Curie RTN
      ``MCnet'' (contract number MRTN-CT-2006-035606) and ``HEPTOOLS''
      (contract number MRTN-CT-2006-035505).}}
\end{center}
\vfill
\begin{center} {\large J.~Alwall$^1$, S.~H\"oche$^2$, F.~Krauss$^2$,
    N.~Lavesson$^3$, L.~L\"onnblad$^3$, F.~Maltoni$^4$,
    M.L.~Mangano$^5$, M.~Moretti$^6$, C.G.~Papadopoulos$^7$,
    F.~Piccinini$^8$, S.~Schumann$^9$, M.~Treccani$^6$, J.~Winter$^9$,
    M.~Worek$^{10,11}$}
\end{center}

\vskip 0.5cm

\begin{center}
$^1$ SLAC, USA; \\ \noindent
$^2$ IPPP, Durham, UK; \\ \noindent
$^3$ Department of Theoretical Physics, Lund University,
   Sweden; \\ \noindent
$^4$ Centre for Particle Physics and Phenomenology (CP3) \\
Universit\'{e} Catholique de Louvain, Belgium; \\ \noindent
$^5$ CERN, Geneva, Switzerland; \\ \noindent
$^6$ Dipartimento di Fisica and INFN, Ferrara, Italy; \\ \noindent
$^7$ Institute of Nuclear Physics, NCSR Demokritos, Athens, Greece; \\ \noindent
$^8$ INFN, Pavia, Italy; \\ \noindent
$^9$ Institut f\"ur Theoretische Physik, TU Dresden,
   Germany; \\ \noindent
$^{10}$ ITP, Karlsruhe University, Karlsruhe,  Germany; \\ \noindent
$^{11}$ Institute of Physics, University of Silesia, Katowice, Poland. \\ \noindent
\end{center}
\vfill
\begin{abstract}
  We compare different procedures for combining fixed-order tree-level
  matrix-element generators with parton showers.  We use the case of
  W-production at the Tevatron and the LHC to compare different
  implementations of the so-called CKKW and MLM schemes using
  different matrix-element generators and different parton cascades.
  We find that although similar results are obtained in all cases,
  there are important differences.
\end{abstract}
\today \hfill
\vfill
\end{titlepage}

\section{Introduction}
One of the most striking features of LHC final states will be the
large number of events with several hard jets.  Final states with 6
jets from $t\bar{t}$\/ decays will have a rate of almost 1~Hz, with
10-100 times more coming from prompt QCD processes.  The immense amount
of available phase space, and the large acceptance of the detectors,
with calorimeters covering a region of almost 10 units of
pseudo-rapidity ($\eta$), will lead to production and identification of
final states with 10 or more jets.  These events will hide or strongly
modify all possible signals of new physics, which involve the chain
decay of heavy coloured particles, such as squarks, gluinos or the
heavier partners of the top, which appear in little-Higgs models.  Being
able to predict their features is therefore essential.

To achieve this, our calculations need to describe as accurately as
possible both the full matrix elements for the underlying hard
processes, as well as the subsequent development of the hard partons
into jets of hadrons.  However, for the complex final-state topologies
we are interested in, no factorization theorem exists to rigorously
separate these two components.  The main obstacle is the existence of 
several hard scales, like the jet transverse energies and di-jet 
invariant masses, which for a generic multi-jet event will span
a wide range.  This makes it difficult to unambiguously separate the
components of the event, which belong to the ``hard process'' (to be
calculated using a multi-parton amplitude) from those developing
during its evolution (described by the parton shower).  A given
$(n+1)$-jet event can be obtained in two ways: from the
collinear/soft-radiation evolution of an appropriate $(n+1)$-parton
final state, or from an $n$-parton configuration where hard,
large-angle emission during its evolution leads to the extra jet.  A
factorization prescription (in this context this is often called a
``matching scheme'' or ``merging scheme'') defines, on an
event-by-event basis, which of the two paths should be followed.  The
primary goal of a merging scheme is therefore to avoid double counting
(by preventing some events to appear twice, once for each path), as
well as dead regions (by ensuring that each configuration is generated
by at least one of the allowed paths).  Furthermore, a good merging
scheme will optimize the choice of the path, using the one, which
guarantees the best possible approximation to a given kinematics.  
It is possible to consider therefore different merging schemes, all
avoiding the double counting and dead regions, but leading to
different results in view of the different ways the calculation is
distributed between the matrix element and the shower evolution.  As in
any factorization scheme, the physics is independent of the separation
between phases only if we have complete control over the perturbative
expansion. Otherwise a residual scheme-dependence is left. Exploring
different merging schemes is therefore crucial to assess the
systematic uncertainties of multi-jet calculations.

In this work we present a comprehensive comparison, for $W$ plus
multijet production, of three merging approaches: the CKKW scheme, the
L\"onnblad scheme, and the MLM scheme. Our investigation is an evolution
and extension of the work in \cite{Mrenna:2003if}, where Mrenna and
Richardson presented implementations of CKKW for \herwig and the
so-called pseudo-shower alternative to CKKW using \pythia, as well as
the results of an approach inspired by the MLM-scheme. Our work
considers the predictions of five different codes, \alpgen, \ariadne,
\helac, \madevent and \sherpa. \alpgen implements the MLM scheme, and
the results shown here are obtained with the \herwig shower; \ariadne
the L\"onnblad scheme; \helac the MLM scheme, but will show results with
the \pythia shower; \madevent uses a variant of the MLM scheme, based on
the CKKW parametrization of the multiparton phase-space; \sherpa,
finally, implements the CKKW scheme. This list of codes therefore covers
a broad spectrum of alternative approaches and, in particular, includes
all the programs used as reference event generators for multijet
production by the Tevatron and LHC experimental collaborations; for
those, we show results relative to publically available versions,
therefore providing valuable information on the systematics involved in
the generation of multijet configurations by the experiments. A
preliminary study, limited to the \alpgen, \ariadne and \sherpa codes,
was presented in~\cite{Hoche:2006ph}.

While \cite{Mrenna:2003if} devoted a large effort to discussing the
internal consistency and validation of the meging schemes, we refer for
these more technical aspects to the papers documenting the individual
implementations of the meging algorithms in the codes we use
\cite{Krauss:2004bs, Krauss:2005nu, Gleisberg:2005qq, Lavesson:2005xu,
Mangano:2006rw}, and we shall limit ourselves here to a short review of
each implementation. We concentrate instead on comparisons among
physical observables, such as cross sections or jet distributions, which
we study for both the Tevatron and the LHC. The main goal is not an
anatomy of the origin of possible differences, but rather the
illustration of their features and their size, to provide the
experimentalists with a quantitative picture of systematics associated
to the use of these codes. We furthermore verify that, with only a few
noteworthy exceptions, the differences among the results of the various
codes are comparable in size with the intrinsic systematics of each
approach, and therefore consistent with a leading-logarithmic level of
accuracy. The quantaties we present correspond to experimental
observables and the differences between the predictions of the various
codes that we present could therefore be resolved by comparing with
data.

We begin the paper with a short review of the merging prescriptions and
of their implementations in the 5 codes. We then introduce the
observables considered for this study, and present detailed numerical
results for both the Tevatron and the LHC. We then provide with an
assessment of the individual systematics of each code, and a general
discusison of our findings.

\section{Merging procedures}
\label{sec:matching-procedures}

In general, the different merging procedures follow a similar
strategy:
\begin{enumerate}
\item A jet measure is defined and all relevant cross sections
  including jets are calculated for the process under consideration.
  I.e.\ for the production of a final state $X$\/ in $pp$-collisions,
  the cross sections for the processes $pp\to X+n\ \mbox{\rm jets}$ with
  $n=0,\,1,\,\dots,\,N=n_{\rm max}$ are evaluated.
\item Hard parton samples are produced with a probability proportional
  to the respective total cross section, in a corresponding kinematic
  configuration following the matrix element.
\item The individual configurations are accepted or rejected with a
  dynamical, kinematics-dependent probability that includes both
  effects of running coupling constants and of Sudakov form factors. In
  case the event is rejected, step 2 is repeated, i.e.\ a new parton
  sample is selected, possibly with a new number of jets.
\item The parton shower is invoked with suitable initial conditions
  for each of the legs. In some cases, like, e.g.\ in the MLM
  procedure described below, this step is performed together with the
  step before, i.e.\ the acceptance/rejection of the jet
  configuration.  In all cases the parton shower is constrained not to
  produce any extra jet; stated in other words: configurations that
  would fall into the realm of matrix elements with a higher jet
  multiplicity are vetoed in the parton shower step.
\end{enumerate}
The merging procedures discussed below differ mainly
\begin{itemize}
\item in the jet definition used in the matrix elements;
\item in the way the acceptance/rejection of jet configurations
  stemming from the matrix element is performed;
\item and in details concerning the starting conditions of and the jet
  vetoing inside the parton showering.
\end{itemize}

\subsection{CKKW}
\label{sec:ckkw}

The merging prescription proposed in \cite{Catani:2001cc,Krauss:2002up} is 
known as the CKKW scheme and has been implemented in the event generator 
\sherpa \cite{Gleisberg:2003xi} in full generality \cite{Schalicke:2005nv}.

In this scheme
\begin{itemize}
\item the separation of the matrix-element and parton-shower domains for 
  different multi-jet processes is achieved through a \kperp-measure 
  \cite{Catani:1991hj,Catani:1992zp,Catani:1993hr}, where $k_{\perp 0}$ 
  denotes the internal separation cut, also called the merging scale;
\item the acceptance/rejection of jet configurations proceeds through
  a reweighting of the matrix elements with analytical Sudakov form
  factors and factors due to different scales in $\alpha_s$;
\item the starting scale for the parton shower evolution of each
  parton is given by the scale where it appeared first;
\item a vetoed parton-shower algorithm is used to guarantee that no
  unwanted hard jets are produced during jet evolution. 
\end{itemize}

In the original paper dealing with $e^+e^-$ annihilations into hadrons,
\cite{Catani:2001cc}, it has been shown explicitly that in this approach 
the dependence on $k_{\perp 0}$ cancels to NLL accuracy.  This can be 
achieved by combining the Sudakov-reweigthed matrix elements with a vetoed
parton shower with angular ordering, subjected to appropriate starting 
conditions.  The algorithm for the case of hadron--hadron collisions 
has been constructed in analogy to the $e^+e^-$ case.  However, it should 
be stressed that it has not been shown that the CKKW algorithm is correct 
at any logarithmic order in this kind of process. 

For hadron-hadron collisions, the internal jet identification of the 
\sherpa-merging approach proceeds through a \kperp-scheme, which defines two 
final-state particles to belong to two different jets, if their relative 
transverse momentum squared
\begin{eqnarray}
  k_{\perp ij}^2 = 2\,\mbox{\rm min}\left\{p_{\perp i},
  \,p_{\perp j}\right\}^2
  \frac{\left[\cosh(\eta_{i}-\eta_{j})-\cos(\phi_{i}-\phi_{j})\right]}
       {D^2}\label{eq:sherpakt}
\end{eqnarray}
is larger than the critical value $k^2_{\perp 0}$. In addition, the
transverse momentum of each jet has to be larger than the merging
scale $k_{\perp 0}$. The magnitude $D$, which is of order $1$, is a
parameter of the jet algorithm \cite{Blazey:2000qt}. In order to
completely rely on matrix elements for jet production allowed by the
external analysis, the internal $D$\/ should be chosen less than or
equal to the $D$-parameter or, in case of a cone-jet algorithm, the
$R$-parameter employed by the external analysis.

The weight attached to the generated matrix elements consists of two
components, a strong-coupling weight and an analytical Sudakov
form-factor weight. For their determination, a \kperp-jet clustering
algorithm guided by only physically allowed parton combinations is
applied on the initial matrix-element configurations. The identified
nodal $k_{\perp}$-values are taken as scales in the strong-coupling
constants and replace the predefined choice in the initial generation.
The Sudakov weight attached to the matrix elements accounts for having
no further radiation resolveable at $k_{\perp 0}$. The NLL-Sudakov
form factors employed, cf.\ \cite{Catani:1991hj}, are defined by
\begin{eqnarray}
  \Delta_q(Q,Q_0) &=&
  \exp\left\{-\int\limits_{Q_0}^{Q} dq \,\Gamma_q(Q,q)\right\} \;,\nonumber\\
  \Delta_g(Q,Q_0) &=&
  \exp\left\{-\int\limits_{Q_0}^{Q} dq
  \left[ \Gamma_g(Q,q) + \Gamma_f(q) \right]\right\} \;,\label{eq:sud}
\end{eqnarray}
where $\Gamma_{q,g,f}$ are the integrated splitting functions $q\to qg$,
$g \to gg$\/ and $g \to q\bar q$, which are given through
\begin{eqnarray}
  \Gamma_q(Q,q) &=& \displaystyle\frac{2 C_F}{\pi}\frac{\alpha_s(q)}{q}
  \left( \ln \frac Q q - \frac 3 4 \right) \;,\\
  \Gamma_g(Q,q) &=& \displaystyle\frac{2 C_A}{\pi}\frac{\alpha_s(q)}{q}
  \left( \ln \frac Q q - \frac{11}{12} \right) \;,\\
  \Gamma_f(q)   &=& \displaystyle\frac{N_f}{3\pi}\frac{\alpha_s(q)}{q}\;.
\end{eqnarray}
They contain the running coupling constant and the two leading,
logarithmically enhanced terms in the limit $Q_0\ll Q$. The single
logarithmic terms $-3/4$ and $-11/12$ may spoil an interpretation of
the NLL-Sudakov form factor as a non-branching probability. Therefore,
$\Gamma(Q,q)$ is cut off at zero, such that $\Delta_{q,g}(Q,Q_0)$ retains 
its property to define the probability for having no emission resolvable
at scale $Q_0$ during the evolution from $Q$ to $Q_0$. These factors
are used to reweight in accordance to the appearance of external
parton lines. A ratio of two Sudakov form factors
$\Delta(Q,Q_0)/\Delta(q,Q_0)$ accounts for the probability of having
no emission resolvable at $Q_0$ during the evolution from $Q$ to $q$.
Hence, it is employed for the reweighting according to internal parton
lines. The lower limit is taken to be $Q_0=k_{\perp 0}$ or
$Q_0=D\,k_{\perp 0}$ for partons that are clustered to a beam or to
another final state parton, respectively.

The sequence of clusterings, stopped after the eventual identification
of a $2\to2$ configuration (the core process), is used to
reweight the matrix element. Moreover, this also gives a shower
history, whereas the $2\to2$ core process defines the starting
conditions for the vetoed shower. For the example of an identified
pure QCD $2\to2$ core process, the four parton lines left as a result
of the completed clustering will start their evolution at the
corresponding hard scale. Subsequently, additional radiation is
emitted from each leg by evolving under the constraint that any
emission harder than the separation cut $k_{\perp 0}$ is vetoed. The
starting scale of each leg is given by the invariant mass of the
mother parton belonging to the identified QCD splitting, through which
the considered parton has been initially formed.

Finally, it should be noted that the algorithm implemented in
\sherpa does the merging of the sequence of processes $pp\to X+n\
\mbox{\rm jets}$ with $n=0,\,1,\,\dots,\,N$\/ fully automatically --
the user is not required to generate the samples separately and mix
them by hand.

\subsection{The Dipole Cascade and CKKW}
\label{sec:dipole-cascade-ckkw}

The merging prescription developed for the dipole cascade in
the \ariadne program\cite{Lonnblad:1992tz} is similar to CKKW, but
differs in the way the shower history is constructed, and in the way
the Sudakov form factors are calculated. Also, since the \ariadne
cascade is ordered in transverse momentum the treatment of starting
scales is simplified.  Before going into details of the merging 
prescription, it is useful to describe some details of the dipole 
cascade, since it is quite different from conventional parton showers.

The dipole model\cite{Gustafson:1988rq,Gustafson:1986db} as
implemented in the \ariadne program is based around iterating
$2\rightarrow 3$ partonic splittings instead of the usual
$1\rightarrow 2$ partonic splittings in a conventional parton shower.
Gluon emission is modeled as coherent radiation from
colour--anti-colour charged parton pairs. This has the advantage of
eg.\ including first order corrections to the matrix elements for
$e^+e^-\to q\bar{q}$\/ in a natural way and it also automatically
includes the coherence effects modeled by angular ordering in
conventional showers.  The process of quark--anti-quark production
does not come in as naturally, but can be
added\cite{Andersson:1990ki}. The emissions in the dipole cascade are
ordered according to an invariant transverse momentum defined as
\begin{equation}
   q_\perp^2 = \frac{s_{12}s_{23}}{s_{123}},\label{eq:arinpt}
\end{equation}
where $s_{ij}$ is the squared invariant mass of parton $i$ and $j$,
with the emitted parton having index 2.

When applied to hadronic collisions, the dipole model does not
separate between initial- and final-state gluon radiation. Instead all
gluon emissions are treated as coming from final-state
dipoles\cite{Andersson:1989gp,Lonnblad:1996ex}.  To be able to extend
the dipole model to hadron collisions, spatially extended coloured
objects are introduced to model the hadron remnants.  Dipoles
involving hadron remnants are treated in a similar manner to the
normal final-state dipoles.  However, since the hadron remnant is
considered to be an extended object, emissions with small wavelength
are suppressed.  This is modeled by only allowing a fraction of the
remnant to take part in the emission. The fraction that is resolved
during the emission is given by
\begin{equation}
   a(q_\perp) = \left(\frac{\mu}{q_\perp}\right)^\alpha,\label{eq:arsup}
\end{equation}
where $\mu$ is the inverse size of the remnant and $\alpha$ is the
dimensionality. These are semi-classical parameters, which have no
correspondence in conventional parton cascades, where instead a
suppression is obtained by ratios of quark densities in the backward
evolution. The main effect is that the dipole cascade allows for
harder gluon emissions in the beam directions, enabling it to describe
properly eg.\ forward jet rates measured at HERA (see eg.\
\cite{Aktas:2005up}).

There are two additional forms of emissions, which need to be included
in the case of hadronic collisions.  One corresponds to an initial
state $g\rightarrow q \bar{q}$\cite{Lonnblad:1995wk}.  This does not
come in naturally in the dipole model, but is added by hand in a way
similar to that of a conventional initial-state parton
shower\cite{Lonnblad:1995wk}. The other corresponds to the
initial-state $q\to gq$ (with the gluon entering into the hard
sub-process), which could be added in a similar way, but this has not
yet been implemented in \ariadne.

When implementing CKKW for the dipole
cascade\cite{Lonnblad:2001iq,Lavesson:2005xu}, the procedure is
slightly different from what has been described above.  Rather than
using the standard \kperp-algorithm to cluster the state produced by
the matrix-element generator, a complete set of intermediate partonic
states, $S_i$, and the corresponding emission scales, $q_{\perp i}$
are constructed, which correspond to a complete dipole shower history.
Hence, for each state produced by the matrix-element generator,
basically the question \textit{how would \ariadne have generated this
  state} is answered. Note, however, that this means that only
coloured particles are clustered, which differs from eg.\ \sherpa,
where also the $W$ and its decay products are involved in the
clustering.

The Sudakov form factors are then introduced using the Sudakov veto
algorithm.  The idea is that we want to reproduce the Sudakov form
factors used in \ariadne. This is done by performing a trial emission
starting from each intermediate state $S_i$ with $q_{\perp i}$ as a
starting scale. If the emitted parton has a $q_\perp$ higher than
$q_{\perp i+1}$ the state is rejected. This correspond to keeping the
state according to the no-emission probability in \ariadne, which is
exactly the Sudakov form factor.

It should be noted that for initial-state showers, there are two
alternative ways of defining the Sudakov form factor. The definition
in eq.~(\ref{eq:sud}) is used in eg.\ \herwig\cite{Corcella:2000bw},
while eg.\ \pythia\cite{Sjostrand:2000wi,Sjostrand:2003wg} uses a form,
which explicitly includes ratios of parton densities. Although
formally equivalent to leading logarithmic accuracy, only the latter
corresponds exactly to a no-emission probability, and this is the one
generated by the Sudakov veto algorithm. This, however, also means
that the constructed emissions in this case need not only be
reweighted by the running \alps as in the standard CKKW procedure
above, but also with ratios of parton densities, which in the case of
gluon emissions correspond to the suppression due to the extended
remnants in eq.~(\ref{eq:arsup}) as explained in more detail in
\cite{Lavesson:2005xu}, where the complete algorithm is presented.

\subsection{The MLM procedure}
\label{sec:mlm-procedure}
\def    \be             {\begin{equation}}
\def    \ee             {\end{equation}}
\def    \ptpart             {\mbox{$p^{\mathrm{part}}_{\perp}$}}
\def    \etmin        {\mbox{$E_\perp^{\mathrm{min}}$}}
\def    \ptmin             {\mbox{$p_{\perp}^{\mathrm{min}}$}}
\def    \etapart             {\mbox{$\eta_{\mathrm{part}}$}}
\def    \etamax         {\mbox{$\eta_{\mathrm{max}}$}}
\def    \as             {\ifmmode \alpha_s \else $\alpha_s$ \fi}
\def    \qzero            {{\mbox{$Q_0$}}}
\def    \qzerosq            {{\mbox{$Q^2_0$}}}
\def \oatwo {\mbox{$ {\cal O} (\alpha_s^2)$}}
\def    \etaclmax       {\mbox{$\eta^{\mathrm clus}_{\mathrm max}$}}
\def    \rmin           {\mbox{$R_{\mathrm{min}}$}}
\def    \rclus          {\mbox{$R_{\mathrm{clus}}$}}
\def    \drpp           {\mbox{$\Delta R_{\mathrm{pp}}$}}
\def    \ptsq           {\mbox{$p_\perp^2$}}
\def    \pt           {\mbox{$p_\perp$}}
\def    \etaclmax       {\mbox{$\eta^{\mathrm{clus}}_{\mathrm{max}}$}}
\def    \etclus         {\mbox{$E^{\mathrm{clus}}_\perp$}}
\def    \gev            {\mbox{$\mathrm{GeV}$}}

The so-called MLM ``matching'' algorithm is described below.

\begin{enumerate}
\item The first step is the generation of
  parton-level configurations for all final-state parton
  multiplicities $n$ up to a given $N$ ($W+N$ partons). They are
  defined by the following kinematical cuts:
  \begin{equation} \label{eq:cuts}
    \ptpart>\ptmin\; , \quad \vert \etapart \vert < \etamax \; , \quad
    \dRjj>\rmin \; ,
  \end{equation}
  where \ptpart\ and \etapart\ are the transverse momentum and
  pseudo-rapidity of the final-state partons, and \dRjj\ is their minimal
  separation in the $(\eta,\phi)$ plane.  The parameters $\ptmin$,
  $\etamax$ and $\rmin$ are called generation parameters, and are the
  same for all $n=1,\,\dots,\,N$.
\item The renormalization scale is set according to the CKKW
  prescription.  The necessary tree branching structure is defined for
  each event, allowing however only for branchings, which are
  consistent with the colour structure of the event, which in \alpgen\
  is extracted from the matrix-element
  calculation~\cite{Caravaglios:1998yr}. For a pair of final-state
  partons $i$\/ and $j$, we use the \kperp-measure defined by
  \begin{equation}
    d_{ij}=\Delta R_{ij}^2\, {\mathrm{min}}(p_{\perp i}^2,p_{\perp j}^2)\; ,
    \label{eq:di1}
  \end{equation}
  where $\Delta R^2_{ij} =\Delta\eta^2_{ij}+\Delta\phi^2_{ij}$, while
  for a pair of initial/final-state partons we have
  \begin{eqnarray}
    d_{ij}=\ptsq,
    \label{eq:di2}
  \end{eqnarray}
  i.e.\ the $\ptsq$ of the final-state one.
\item The \kperp-value at each vertex is used as a scale for the
  relative power of $\alpha_s$.  The factorization scale for the
  parton densities is given by the hard scale of the process,
  $\qzerosq=m_W^2+p_{\perp W}^2$.  It may happen that the clustering
  process stops before the lowest-order configuration is reached. This
  is the case, e.g., for an event like $u \bar{u} \to W c \bar{s} g$.
  Flavour conservation allows only the gluon to be clustered, since $u
  \bar{u}\to W c \bar{s}$ is a LO process, first appearing at \oatwo.
  In such cases, the hard scale \qzero\ is adopted for all powers of
  \alps corresponding to the non-merged clusters.
\item Events are then showered, using \pythia or \herwig. The
  evolution for each parton starts at the scale determined by the
  default \pythia and \herwig\ algorithms on the basis of the
  kinematics and colour connections of the event. The upper veto
  cutoff to the shower evolution is given by the hard scale of the
  process, \qzero. After evolution, a jet cone algorithm is applied to
  the partons produced in the perturbative phase of the shower.  Jets
  are defined by a cone size \rclus, a minimum transverse energy
  \etclus\ and a maximum pseudo-rapidity $\etaclmax$.  These
  parameters are called matching parameters, and should be kept the
  same for all samples $n=0,\,1,\,\dots,\,N$. These jets provide the
  starting point for the matching procedure, described in the next
  bullet.  In the default implementation, we take $\rclus=\rmin$,
  $\etaclmax=\etamax$ and $\etclus=\ptmin+\mathrm{max}(5\; \gev,0.2
  \times \ptmin)$, but these can be varied as part of the systematics
  assessment. To ensure a complete coverage of phase space, however,
  it is necessary that $\rclus \ge \rmin$, $\etaclmax\le\etamax$ and
  $\etclus \ge \ptmin$.
\item Starting from the hardest parton, the jet, which is closest to
  it in $(\eta,\phi)$ is selected.  If the distance between the parton
  and the jet centroid is smaller than $1.5\times \rclus$, we say that
  the parton and the jet {\em match}.  The matched jet is removed from
  the list of jets, and the matching test for subsequent partons is
  performed. The event is fully matched if each parton matches to a
  jet.  Events, which do not match, are rejected.  A typical example
  is when two partons are so close that they cannot generate
  independent jets, and therefore cannot match.  Another example is
  when a parton is too soft to generate its own jet, again failing
  matching.
\item Events from the parton samples with $n<N$, which survive
  matching, are then required not to have extra jets. If they do, they
  are rejected, a suppression, which replaces the Sudakov reweighting
  used in the CKKW approach. This prevents the double counting of
  events, which will be present in, and more accurately described by,
  the $n+1$ sample. In the case of $n=N$, events with extra jets can
  be kept since they will not be generated by samples with higher $n$.
  Nevertheless, to avoid double counting, we require that their
  transverse momentum be smaller than that of the softest of the
  matched jets.
\end{enumerate}
When all the resulting samples from $n=0,\,\dots,\,N$\/ are combined,
we obtain an inclusive $W+$jets sample.  The harder the threshold for
the energy of the jets used in the matching, \etclus, the fewer the
events rejected by the extra-jet veto (i.e.\ smaller Sudakov
suppression), with a bigger role given to the shower approximation in
the production of jets. Using lower thresholds would instead enhance
the role of the matrix elements even at lower $E_\perp$, and lead to
larger Sudakov suppression, reducing the role played by the shower in
generating jets. The matching/rejection algorithm ensures that these two
components balance each other.  This algorithm is encoded in the
\alpgen\ generator~\cite{Mangano:2001xp,Mangano:2002ea}, where evolution
with both \herwig\ and \pythia\ are enabled.  However, in the framework
of this study, the parton shower evolution has been performed by
\herwig.

\subsection{The \protect\madevent\ approach}
\label{sec:mad-procedure}
The approach used in \madgraph/\madevent
\cite{Stelzer:1994ta,Maltoni:2002qb} is based on the MLM prescription,
but uses a different jet algorithm for defining the scales in \alps
and for the jet matching. The phase-space separation between the
different multi-jet processes is achieved using the \kperp-measure as
in \sherpa (eq.~(\ref{eq:sherpakt}) with $D=1$), while the Sudakov
reweighting is performed by rejecting showered events that do not
match to the parton-level jets, as in \alpgen.  This approach allows
more direct comparisons with \sherpa, including the effects of
changing the \kperp-cutoff scale.  The details of the procedure are as
follows.

Matrix-element multi-parton events are produced using
\madgraph/\madevent version 4.1 \cite{Alwall:2007st}, with a cutoff
$Q_\mathrm{min}^\mathrm{ME}$ in clustered \kperp. The multi-parton
state from the matrix-element calculation is clustered according to
the \kperp-algorithm, but allowing only clusterings that are
compatible with the Feynman diagrams of the process, which are
provided to \madevent by \madgraph. The factorization scale, i.e., the
scale used in the parton densities, is taken to be the clustering
momentum in the last $2\to2$ clustering (the ``central process''),
usually corresponding to the transverse mass, $m_\perp$, of the $W$\/
boson. The \kperp-scales of the QCD clustering nodes are used as
scales in the calculation of the various powers of \alps.

As in the \alpgen procedure, no Sudakov reweighting is performed.
Instead, the virtuality-ordered shower of \pythia 6.4
\cite{Sjostrand:2006za} is used to shower the event, with the starting
scale of the shower set to the factorization scale.
The showered (but not yet hadronized) event is then clustered to jets
using the \kperp-algorithm with a jet measure cutoff
$Q_\mathrm{min}^\mathrm{jet} > Q_\mathrm{min}^\mathrm{ME}$, and the
matrix-element partons are matched to the resulting jets, in a way,
which differs from the standard MLM procedure. A parton is considered
to be matched to the closest jet if the jet measure
$Q(\mathrm{parton},\mathrm{jet})$ is smaller than the cutoff
$Q_\mathrm{min}^\mathrm{jet}$.  Events where not all partons are
matched to jets are rejected. For events with parton multiplicity
smaller than the highest multiplicity, the number of jets must be
equal to the number of partons. For events with the highest
multiplicity, $N$ jets are
reconstructed, and partons are considered to be matched if $Q(\mathrm{parton},
\mathrm{jet}) < Q^{\mathrm{parton}}_{N}$, the smallest $k_\perp$-measure
in the matrix-element event. This means that extra jets below
$Q^{\mathrm{parton}}_{N}$ are allowed, similarly to the Sherpa
treatment.


Note that also the standard MLM scheme with cone jets is implemented
as an alternative in \madevent and its \pythia interface.

\subsection{\protect\helac\ implementation of the MLM procedure}
\label{sec:helac-procedure} 
In \helac \cite{Kanaki:2000ey,Papadopoulos:2005ky} we have implemented
the MLM procedure as described above, see section~\ref{sec:mlm-procedure}.
\helac generates events for all possible processes at hadron and
lepton colliders within the Standard Model and has been successfully
tested with up to 10 particles in the final state
\cite{Papadopoulos:2005ky,Gianotti:2002xx,Gleisberg:2003bi}.

The partons from the matrix-element calculation are matched to the
jets constructed after the parton showering. The parton-level events
are generated with a minimum $p_{\perp \rm min}$ threshold for the
partons, $p_{\perp j}>p_{\perp \rm min}$, a minimum parton separation,
$\Delta R_{jj}>R_{\rm min}$, and a maximum pseudo-rapidity,
$|\eta_j|<\eta_{\rm max}$.
In order to extract the necessary information used by the
\kperp-reweighting, initial- and final-state partons are clustered
backwards as described in section~\ref{sec:mlm-procedure}, where again
the colour flow information extracted from the matrix-element
calculation is used as a constraint on the allowed clusterings. 
The \kperp-measure, $d_{ij}$, for pairs of outgoing partons is given by
equation (\ref{eq:di1}) and for pairs of partons where one is incoming
and one is outgoing by equation (\ref{eq:di2}). If two outgoing partons
are clustered, i.e.\ $d_{ij}$ is minimal, the resulting parton is again
an outgoing parton with $p=p_i+p_j$ and adjusted colour flow. In the
case when incoming and outgoing partons are clustered, the new parton is
incoming and its momentum is $p=p_j-p_i$. As a
result we obtain a chain of $d$-values. For every node, a factor
of $\alpha_s(d_{\rm node})/\alpha_s(Q^2_0)$ is multiplied
into the weight of the event. For the unclustered vertices as well as
for the scale used in the parton density functions, the hard scale 
of the process $Q_0^2=m^2_W+p^2_{\perp W}$ is used. No Sudakov
reweighting is applied. The sample of events output, which is in the
latest Les Houches event file format~\cite{Alwall:2006yp}, is read by
the interface to \pythia version 6.4 \cite{Sjostrand:2006za}, where
the virtuality-ordered parton shower is constructed. For each event, a
cone jet-algorithm is applied to all partons resulting from the shower
evolution. The resulting jets are defined by $E_{\perp \rm min}^{\rm
clus}$, $\eta_{\rm max}^{\rm clus}$ and by a jet cone size $R_{\rm
clus}$. The parton from the parton-level event is then associated to
one of the constructed jets. Starting from the parton with the highest
$p_\perp$ we select the closest jet ($1.5\,\times\,R_{\rm clus}$) in the
pseudo-rapidity/azimuthal-angle space. All subsequent partons are
matched iteratively to jets. If this is impossible, the event is
rejected. Additionally, for $n<N$,  matched events with the number of
jets greater than $n$\/ are rejected, whereas for $n=N$, i.e.\ the
highest multiplicity (in this study, $N=4$), events with extra jets
are kept, only if they are softer than the $N$\/ matched jets. This
procedure provides the complete inclusive sample.

\section{General properties of the event generation for the study}
\label{sec:genprop}

We present in the following sections some concrete examples. We
concentrate on the case of $W$+multi-jet production, which is one of
the most studied final states because of its important role as a
background to top quark studies at the Tevatron. At the LHC, $W$+jets,
as well as the similar $Z$+jets processes, will provide the main
irreducible backgrounds to signals such as multi-jet plus missing
transverse energy, typical of Supersymmetry and of other
manifestations of new physics. The understanding of $W$+multi-jet
production at the Tevatron is therefore an essential step towards the
validation and tuning of the tools presented here, prior to their
utilization at the LHC.

The CDF and D\O\ experiments at the Tevatron collider  have
reported cross-section measurements for $W$+multijet final states, both
from Run~I \cite{Abe:1993si,Abe:1997eva,Affolder:2000mz,Abachi:1995jf}
and, in preliminary form, from Run~II ~\cite{Messina:2006zz}. The Run~I
results typically refer to detector-level quantities, and a comparison
with theoretical predictions requires to process the generated events
through a detector simulation. These tests were performed in the context
of the quoted analyses, using the LO calculations available at the time,
showing a good agreemnt within the large statistical, systematic and
theoretical uncertainties. The preliminary CDF result from Run~II
~\cite{Messina:2006zz} is instead corrected for all detector effects,
and expressed in terms of {\em true} jet energies. In this form it is
therefore suitable for direct comparison with theory predictions.
Measurements of $Z$+multijet rates are also crucial, but suffer from
lower statistics w.r.t.\ the $W$ case. A Run~II measurement of jet
$p_\perp$ spectra in $Z$+multijet events from D\O\ has been
compared to the predictions of \sherpa in ref.~\cite{Abazov:2006gs},
showing again a very good agreement. Preliminary CDF results on the
spectra of the first and second jet in $Z$+jet events have been compared
against parton-level NLO results~\cite{CDFZjet}. For both the $W$ and
$Z$ cases, the forthcoming analyses of the high-statistics sample now
available at the Tevatron will provide valuable inputs for more
quantitative analyses of the codes presented here.

For each of the codes, we calculated a large set of observables,
addressing inclusive properties of the events (transverse momentum
spectrum of the $W$\/ and of leading jets) as well as geometric
correlations between the jets. What we present and discuss here is a
subset of our studies, which illustrates the main features of the
comparison between the different codes and of their own systematics. A
preliminary account of these results, limited to the \alpgen, \ariadne
and \sherpa codes, was presented in~\cite{Hoche:2006ph}.  More
complete studies of the systematics of each individual code have been
~\cite{Krauss:2004bs,Krauss:2005nu,Gleisberg:2005qq,Lavesson:2005xu,Mangano:2006rw}
or will be presented elsewhere by the respective authors.

The existence in each of the codes of parameters specifying the
details of the merging algorithms presents an opportunity
to tune each code so as to best describe the data. This tuning should
be seen as a prerequisite for a quantitative study of the overall
theoretical systematics: after the tuning is performed on a given set
of final states (e.g.\ the $W$+jets considered here), the systematics
for other observables or for the extrapolation to the LHC can be
obtained by comparing the difference in extrapolation between the
various codes. Here it would be advantageous if future analysis of
Tevatron data would provide us with spectra corrected for detector
effects in a fashion suitable for a direct comparison against
theoretical predictions.

The following two sections present results for the Tevatron
($p\bar{p}$ collisions at 1.96~TeV) and for the LHC ($pp$ at
14~TeV). The elements of the analysis common to all codes are the
following:
\begin{itemize}
\item {\it Event samples.} Tevatron results refer to the combination
  of $W^+$ and $W^-$ bosons, while at the LHC only $W^+$ are
  considered. All codes have generated parton-level samples according
  to matrix elements with up to 4 final-state partons, i.e.\ $N=4$.
  Partons are restricted to the light-flavour sector and are taken to
  be massless. The Yukawa couplings of the quarks are neglected. The
  PDF set CTEQ6L has been used with $\alpha_s(m_Z)=0.118$. Further
  standard-model parameters used were: $m_W=80.419$~GeV,
  $\Gamma_W=2.048$~GeV, $m_Z=91.188$~GeV, $\Gamma_Z=2.446$~GeV, the
  Fermi constant $G_\mu=1.16639\cdot10^{-2}\mbox{~GeV}^{-2}$,
  $\sin^2\theta_W=0.2222$ and $\alpha_{\mbox{\scriptsize
      EM}}=1/132.51$.
\item {\it Jet definitions.} Jets were defined using Paige's {\tt
    GETJET} cone-clustering algorithm, with a calorimeter segmentation
  of ($\Delta \eta$, $\Delta\phi$) = (0.1,$6^\circ$) extended over the
  range $\vert\eta\vert<2.5$ ($\vert\eta\vert<5$), and cone size of
  0.7 (0.4) for the Tevatron (LHC).  At the Tevatron (LHC) we require
  jets with $E_\perp>10\ (20)$~GeV, and pseudo-rapidity
  $\vert\eta\vert<2\ (4.5)$. For the analysis of the differential
    jet rates denoted as $d_i$, the Tevatron Run~II \kperp-algorithm
  \cite{Blazey:2000qt}\footnote{More precisely, we used the
    implementation in the {\tt ktclus} package \cite{ktclus} ({\tt
      IMODE=5}, or {\tt 4211}).} was applied to all final-state
  particles fulfilling $\vert\eta\vert<2.5\ (5)$. The \kperp-measure
  used in the algorihtm is given by equations (\ref{eq:di1}) and
  (\ref{eq:di2}).
\end{itemize}
In all cases, except the $d_i$ plots, the analysis is done at the hadron
level, but without including the underlying event. The $d_i$ plots were
done to check the details of the merging and are therefore done at
parton level to avoid any smearing effects from hadronization. For all
codes, the systematic uncertainties are investigated by varying the
merging scale and by varying the scale in \alps and, for some codes, in
the parton density functions. For \alpgen and \helac, the scale in \alps
has been varied only in the \alps-reweighting of the matrix elements,
while for the others the scale was also varied in the parton cascade.
Note that varying the scale in the final-state parton showers will spoil
the tuning done to LEP data for the cascades. A consistent way of
testing the scale variations would require retuning of hadronization
parameters. However, we do not expect a strong dependence on the
hadronization parameters in the observables we consider, and no attempt
to retune has been made.

The parameter choices specific to the individual codes are as follows:
\begin{itemize}

\item \alpgen: The parton-level matrix elements were generated with
  \alpgen~\cite{Mangano:2001xp,Mangano:2002ea} and the subsequent
  evolution used the \herwig parton shower according to the MLM
  procedure. Version 6.510 of \herwig was used, with its default
  shower and hadronization parameters.  The {\em default} results for
  the Tevatron (LHC) were obtained using parton-level cuts (see
  eq.(\ref{eq:cuts})) of $\ptmin=8\ (15)$ GeV, $ \etamax=2.5\ (5)$,
  $\rmin=0.7\ (0.4)$ and matching defined by $\etclus=10\ (20)$~GeV,
  $\etaclmax=\etamax$ and $\rclus=\rmin$. The variations used in the
  assessment of the systematics cover:
  \begin{itemize}
  \item different thresholds for the definition of jets used in the
    matching: $\etclus=20$ and 30~GeV for the Tevatron, and
    $\etclus=30$ and 40~GeV for the LHC. These thresholds were
    applied to the partonic samples produced with the default
    generation cuts, as well as to partonic samples produced with
    higher $\ptmin$ values. No difference was observed in the results,
    aside from an obviously better generation efficiency in the latter
    case. In the following studies of the systematics, the two threshold
    settings will be referred to as \alpgen parameter sets ALptX,
    where X labels the value of the threshold. Studies with different
    values of \rclus\ and \rmin\ were also performed, leading to
    marginal changes, which will not be documented here.
  \item different renormalization scales at the vertices of the clustering
    tree: $\mu=\mu_0/2$ and $\mu=2\,\mu_0$, where $\mu_0$ is the default
    \kperp-value. In the following studies of
    the systematics, these two
    settings will be referred to as \alpgen parameter sets ALscL (for
    ``Low'')  and ALscH (for ``High'').
  \end{itemize}
The publicly available version V2.10 of the code was used to generate
all the \alpgen results.

\item \ariadne: The parton-level matrix elements were generated with
  \madevent and the subsequent evolution used the dipole shower in
  \ariadne according to the procedure outlined in
  section~\ref{sec:dipole-cascade-ckkw}. Hadronization was performed by
  \pythia.

  For the {\em default} results at the Tevatron (LHC) the parton-level
  cuts were $p_{\perp \rm min}=10\ (20)$, $R_{jj}<0.5\ (0.4)$ and, in
  addition, a cut on the maximum pseudo-rapidity of jets,
  $\eta_{j\rm max}=2.5\ (5.0)$. The variations used in the assessment
  of the systematics cover:
  \begin{itemize}
  \item different values of the merging scales $p_{\perp \rm min}=20$ and
    30~GeV for the Tevatron (30 and 40~GeV for the LHC). In the
    following studies of the systematics, these two settings will be
    referred to as \ariadne parameter sets ARptX.
  \item a change of the soft suppression parameters in
    eq.~(\ref{eq:arsup}) from the default values of
    $\mu=0.6$~GeV and $\alpha=1$, to $\mu=0.6$~GeV and $\alpha=1.5$
    (taken from a tuning to HERA data \cite{Brook:1995nn}). This setting
    will be referred to as ARs.
  \item different values of the scale in \alps: $\mu=\mu_0/2$
    and $\mu=2\,\mu_0$ were used (ARscL and ARscH). This scale change
    was used in \alps evaluations in the program.
  \end{itemize}

\item \helac: The parton-level matrix elements were generated with
  \helac~\cite{Kanaki:2000ey,Papadopoulos:2005ky} and the phase space
  generation is performed by \phegas~\cite{Papadopoulos:2000tt}. The
  subsequent evolution used the default virtuality-ordered shower in
  \pythia 6.4 \cite{Sjostrand:2006za} according to the
  MLM procedure. Hadronization was performed by \pythia.

  In the present study, $e^+\nu_e+n$ jets and $e^-\bar{\nu}_e+n$ jets
  samples with $n=0,\,\dots,\,4$ have been generated for Tevatron,
  while for LHC predictions only $e^+\nu_e+n$ jets final states have been
  considered. The number of subprocesses (i.e.\ $u\bar{d}\to e^+ \nu_e
  u \bar{u} g g$ is one for the $W^+ + 4$ jets) in those cases
  is 4, 12, 94, 158 and 620 for $n=0,\,1,\,2,\,3,\,4$ respectively,
  with the number of quark flavours being 4/5 for the initial/final
  states.

  The {\em default} results for the Tevatron (LHC) were obtained using
  parton-level cuts of $p_{\perp \rm min}=8\ (15)$ GeV, $\eta_{\rm
  max}=2.5\ (5)$, $R_{\rm min}=0.7\ (0.4)$ and matching defined by
  $E_{\perp \rm min}^{\rm clus}=10\ (20)$~GeV, $\eta_{\rm max}^{\rm
  clus}=2\ (4.5)$ and $R_{\rm min}^{\rm clus}=0.7\ (0.4)$. The
  variations used in the assessment of the systematics cover:
  \begin{itemize}
  \item different thresholds for the definition of jets used in the
    matching: $E_{\perp \rm min}^{\rm clus}=30$~GeV for the Tevatron, and
    $E_{\perp \rm min}^{\rm clus}=40$~GeV for the LHC. In the following
    studies of the systematics, these two settings will be referred to
    as \helac parameter sets HELptX, where X labels the value of the
    threshold.
  \item different renormalization scales at the vertices of the
    clustering tree: $\mu=\mu_0/2$ and $\mu=2\,\mu_0$, where $\mu_0$ is
    the default \kperp-value. In the following studies of the
    systematics, these two settings will be referred to as \helac
    parameter sets HELscL and HELscH.
 \end{itemize}

\item \madevent: The parton-level matrix elements were generated with
  \madevent and the subsequent evolution used the \pythia shower
  according to the modified MLM procedure in
  section~\ref{sec:mad-procedure}. Hadronization was performed by \pythia.

  For the {\em default} results at the Tevatron (LHC) the value of the
  merging scale has been chosen to $k_{\perp 0}=10\ (20)$~GeV. The
  variations used in the assessment of the systematics cover:
  \begin{itemize}
  \item different values of the merging scale $k_{\perp 0}=20$ and
    30~GeV for the Tevatron, and $k_{\perp 0}=30$ and 40~GeV for the
    LHC. In the following studies of the systematics, these two
    settings will be referred to as \madevent parameter sets MEktX.
  \item different values of the scales used in the evaluation of
    \alps, in both the matrix element generation and the parton
    shower: $\mu=\mu_0/2$ and $\mu=2\,\mu_0$, where $\mu_0$ is the
    default \kperp-value.  These two settings will be referred to as
    \madevent parameter sets MEscL and MEscH.
  \end{itemize}

\item \sherpa: The parton-level matrix elements used within \sherpa
  have been obtained from the internal matrix-element generator
  \amegic \cite{Krauss:2001iv}. Parton showering has been conducted by
  \apacic \cite{Kuhn:2000dk,Krauss:2005re} whereas the combination of
  the matrix elements with this parton shower has been accomplished
  according to the CKKW procedure\footnote{
    Beyond the comparison presented here, \sherpa predictions for
    $W$+multi-jets have already been validated and studied for
    Tevatron and LHC energies in \cite{Krauss:2004bs,Krauss:2005nu}.
    Results for the production of pairs of $W$-bosons have been
    presented in \cite{Gleisberg:2005qq}.}.
  The hadronization of the shower configurations has been performed by
  \pythia 6.214, which has been made available through an internal
  interface.

  For the {\em default} Tevatron (LHC) predictions, the value of the
  merging scale has been chosen to $k_{\perp 0}=10\ (20)$~GeV. All
  \sherpa predictions for the Tevatron (LHC) have been obtained by
  setting the internally used $D$-parameter (cf.\
  eq.~(\ref{eq:sherpakt}) in section~\ref{sec:ckkw}) through
  $D=0.7\ (0.4)$. Note that, these two choices directly determine the
  generation of the matrix elements in \sherpa. The variations used in
  the assessment of the systematics cover:
  \begin{itemize}
  \item first, different choices of the merging scale $k_{\perp 0}$.
    Values of 20 and 30~GeV, and 30 and 40~GeV have been
    used for the Tevatron and the LHC case, respectively. In the
    following studies of the systematics, these settings will be
    referred to as \sherpa parameter sets SHktX where X labels the
    value of the internal jet scale.
  \item and, second, different values of the scales used in any
    evaluation of the \alps\ {\it and}\/ the parton distribution
    functions\footnote{For example, the analytical Sudakov form
      factors used in the matrix-element reweighting hence vary owing
      to their intrinsic \alps-coupling dependence.}.
    Two cases have been considered, $\mu=\mu_0/2$ and $\mu=2\,\mu_0$.
    The choice of the merging scale is as in the default run, where
    $\mu_0$ denotes the corresponding \kperp-values. In the subsequent
    studies of the systematics these two cases are referred to as
    \sherpa parameter sets SHscL and SHscH. It should be stressed that
    these scale variations have been applied in a very comprehensive
    manner, i.e.\ in both the matrix-element and parton-showering
    phase of the event generation.
  \end{itemize}
  All \sherpa results presented in this comparison have been obtained
  with the publicly available version 1.0.10.

\end{itemize}

\section{Tevatron Studies}
\label{sec:tevatron-results}

\subsection{Event rates}
\label{sec:tevrates}

\begin{figure}
\begin{center}
\includegraphics[width=0.8\textwidth,clip]{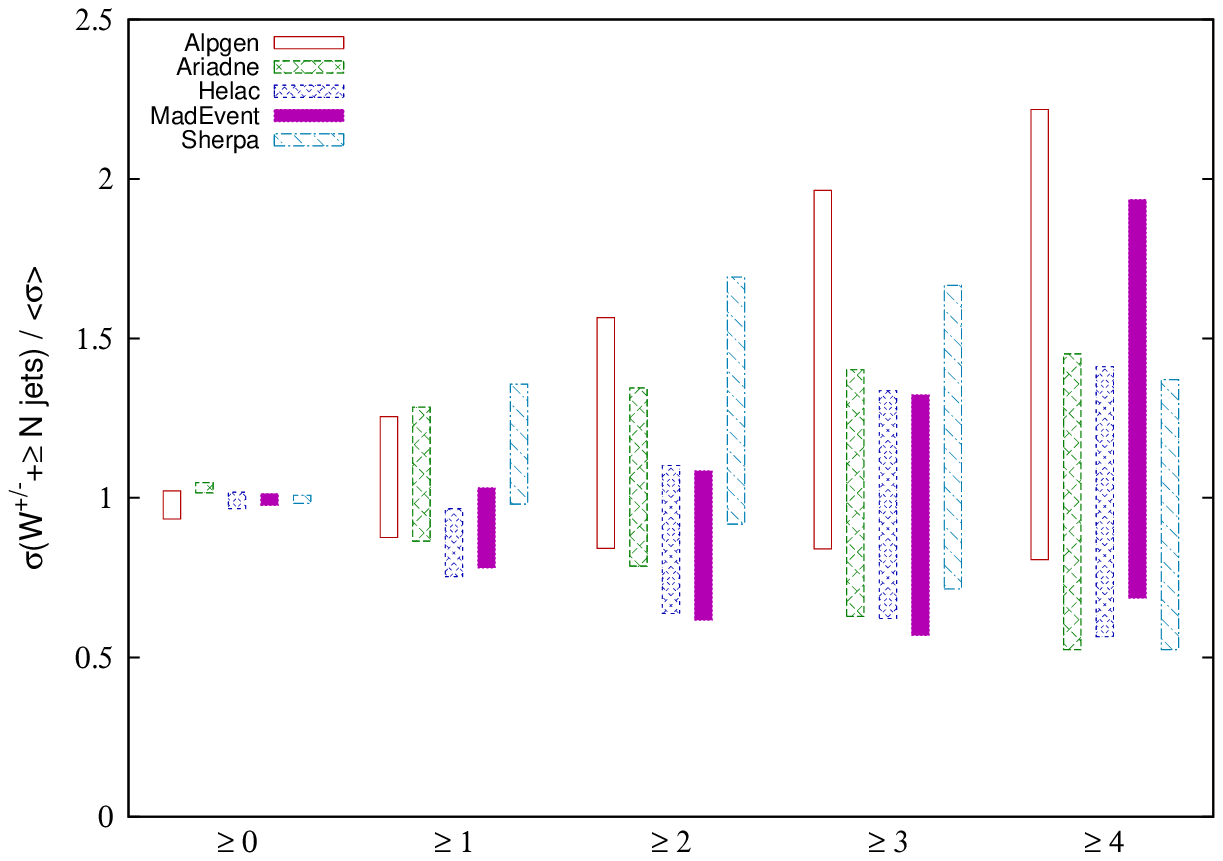}
\end{center}
\vskip -0.4cm
\ccaption{}{\label{fig:rates-tev}Range of variation for the Tevatron
  cross-section rates of the five codes, normalized to the average
  value of the default settings for all codes in each multiplicity
  bin.}
\end{figure}

We present here the comparison among inclusive jet rates. These are
shown in table~\ref{tab:tevrates}. For each code, in addition to the
default numbers, we present the results of the various individual
alternative choices used to assess the systematics uncertainty. 
In table~\ref{tab:tevratios} we show the ``additional jet fractions'',
namely the rates $\sigma(W+n+1\ \mathrm{jets})/\sigma(W+n\
\mathrm{jets})$, once again covering all systematic sets of all codes.
\Fig{fig:rates-tev}, finally, shows graphically the
cross-section systematic ranges: for each multiplicity, we normalize
the rates to the average of the default values of all the codes. 

It should be noted that the scale changes in all codes lead to the
largest rate variations. This is reflected in the growing size of the
uncertainty with larger multiplicities, a consequence of the higher
powers of \alps. A more detailed discussion on the effects of the scale
changes can be found in section~\ref{sec:syst}. Furthermore we note that
the systematic ranges of all codes have regions of overlap.

\begin{table}
\begin{center}
\begin{tabular}{|l|rrrrr|}
\hline
Code & $\sigma$[tot] & $\sigma$[$\ge 1$ jet] & $\sigma$[$\ge 2$ jet]
&$\sigma$[$\ge 3$ jet] &$\sigma$[$\ge 4$ jet]  \\
\hline
{\bf \alpgen, def}
& {\bf 1933} & {\bf 444} & {\bf 97.1} & {\bf 18.9} & {\bf 3.2} \\
ALpt20        & 1988 & 482 & 87.2 & 15.5 & 2.8\\
ALpt30        & 2000 & 491 & 82.9 & 12.8 & 2.1 \\
ALscL       & 2035 & 540 & 135  & 29.7 & 5.5  \\
ALscH        & 1860 & 377 & 72.6 & 12.7 & 2.0 \\
\hline
{\bf \ariadne, def}
& {\bf 2066} & {\bf 477} & {\bf 87.3} & {\bf 13.9} & {\bf 2.0} \\
ARpt20           & 2038 & 459 & 76.6 & 12.8 & 1.9 \\
ARpt30          & 2023 & 446 & 67.9 & 11.3 & 1.7 \\
ARscL           & 2087 & 553 & 116 & 21.2 & 3.6\\
ARscH           & 2051 & 419 & 67.8 & 9.5 & 1.3\\
ARs             & 2073 & 372 & 80.6 & 13.2 & 2.0\\

\hline
{\bf \helac, def}
& {\bf 1960} & {\bf  356} & {\bf 70.8} & {\bf 13.6} &{\bf 2.4} \\
HELpt30 & 1993 & 373 & 68.0 & 12.5  & 2.4 \\
HELscL & 2028 & 416 & 95.0 & 20.2 & 3.5 \\
HELscH &  1925 & 324  & 55.1   & 9.4  & 1.4 \\
\hline

{\bf \madevent, def}
& {\bf 2013} & {\bf 381} & {\bf 69.2} & {\bf 12.6} &{\bf 2.8} \\
MEkt20 & 2018 & 375 & 66.7 & 13.3 & 2.7\\
MEkt30 & 2017 & 361 & 64.8 & 11.1 & 2.0\\
MEscL & 2013 & 444 & 93.6 & 20.0 & 4.8\\
MEscH & 1944 & 336 & 53.2 & 8.6 & 1.7\\
\hline
{\bf \sherpa, def}
& {\bf 1987} & {\bf 494} & {\bf 107} & {\bf 16.6} &{\bf 2.0} \\
SHkt20 & 1968 & 465 & 85.1 & 12.4 & 1.5 \\
SHkt30 & 1982 & 461 & 79.2 & 10.8  & 1.3 \\
SHscL & 1957 & 584 & 146  & 25.2 & 3.4 \\
SHscH & 2008 & 422 & 79.8 & 11.2 & 1.3 \\
\hline
\hline
\end{tabular}
\ccaption{}{\label{tab:tevrates} Cross sections (in pb) for the
  inclusive jet rates at the Tevatron, according to the default and
  alternative settings of the various codes.}
\end{center}
\end{table}

\begin{table}
\begin{center}
\begin{tabular}{|lrrrr|}
\hline
Code & $\sigma^{[\ge 1]}$/$\sigma^{[\mathrm tot]}$ &
       $\sigma^{[\ge 2]}$/$\sigma^{[\ge 1]}$ &
       $\sigma^{[\ge 3]}$/$\sigma^{[\ge 2]}$ &
       $\sigma^{[\ge 4]}$/$\sigma^{[\ge 3]}$ \\
\hline
{\bf \alpgen, def } & {\bf 0.23} & {\bf 0.22} &
{\bf 0.19} & {\bf 0.17} \\
ALpt20          & 0.24 & 0.18 & 0.18 & 0.18 \\
ALpt30          & 0.25 & 0.17 & 0.15 & 0.16\\
ALscL         & 0.27 & 0.25 & 0.22 & 0.19 \\
ALscH          & 0.20 & 0.19 & 0.17 & 0.16 \\
\hline
{\bf \ariadne, def} & {\bf 0.23} & {\bf 0.18} &
{\bf 0.16} & {\bf 0.15} \\
ARpt20         & 0.23 & 0.17 & 0.17 & 0.15 \\
ARpt30         & 0.22 & 0.15 & 0.16 & 0.16 \\
ARscL             & 0.26 & 0.21 & 0.18 & 0.17 \\
ARscH             & 0.20 & 0.16 & 0.14 & 0.14\\
ARs         & 0.18 & 0.22 & 0.16 & 0.15 \\
\hline
{\bf \helac, def} &  {\bf 0.18} &  {\bf 0.20} &
{\bf 0.19} &  {\bf 0.18} \\
HELpt30 &  0.19 &  0.19 &  0.18 &  0.19 \\
HELscL  &  0.21 &  0.23 &  0.21 &  0.17 \\
HELscH  &   0.17 &  0.17 &  0.17 &  0.15 \\
\hline
{\bf \madevent, def} & {\bf 0.19} & {\bf 0.18} &
{\bf 0.18} & {\bf 0.22}\\
MEkt20            & 0.19 & 0.18 & 0.20 & 0.20 \\
MEkt30            & 0.18 & 0.18 & 0.17 & 0.18 \\
MEscL            & 0.22 & 0.21 & 0.21 & 0.24 \\
MEscH            & 0.17 & 0.16 & 0.16 & 0.20 \\
                \hline
{\bf \sherpa, def} &   {\bf 0.25} &  {\bf 0.22} &
{\bf 0.16} &  {\bf 0.12} \\
SHkt20 & 0.24 &  0.18 &  0.15 &  0.12 \\
SHkt30 &  0.23 &  0.17 &  0.14 &  0.12\\
SHscL &  0.30 &  0.25 &  0.17 &  0.13\\
SHscH &  0.21 &  0.19 &  0.14 &  0.12\\ 
\hline
\hline
\end{tabular}
\ccaption{}{\label{tab:tevratios} Cross-section ratios for $(n+1)/n$\/
  inclusive jet rates at the Tevatron, according to the default and
  alternative settings of the various codes.}
\end{center}
\end{table}

\subsection{Kinematical distributions}
\label{sec:tevdists}

\begin{figure}
  \begin{center}
    \includegraphics[width=0.92\textwidth,clip]{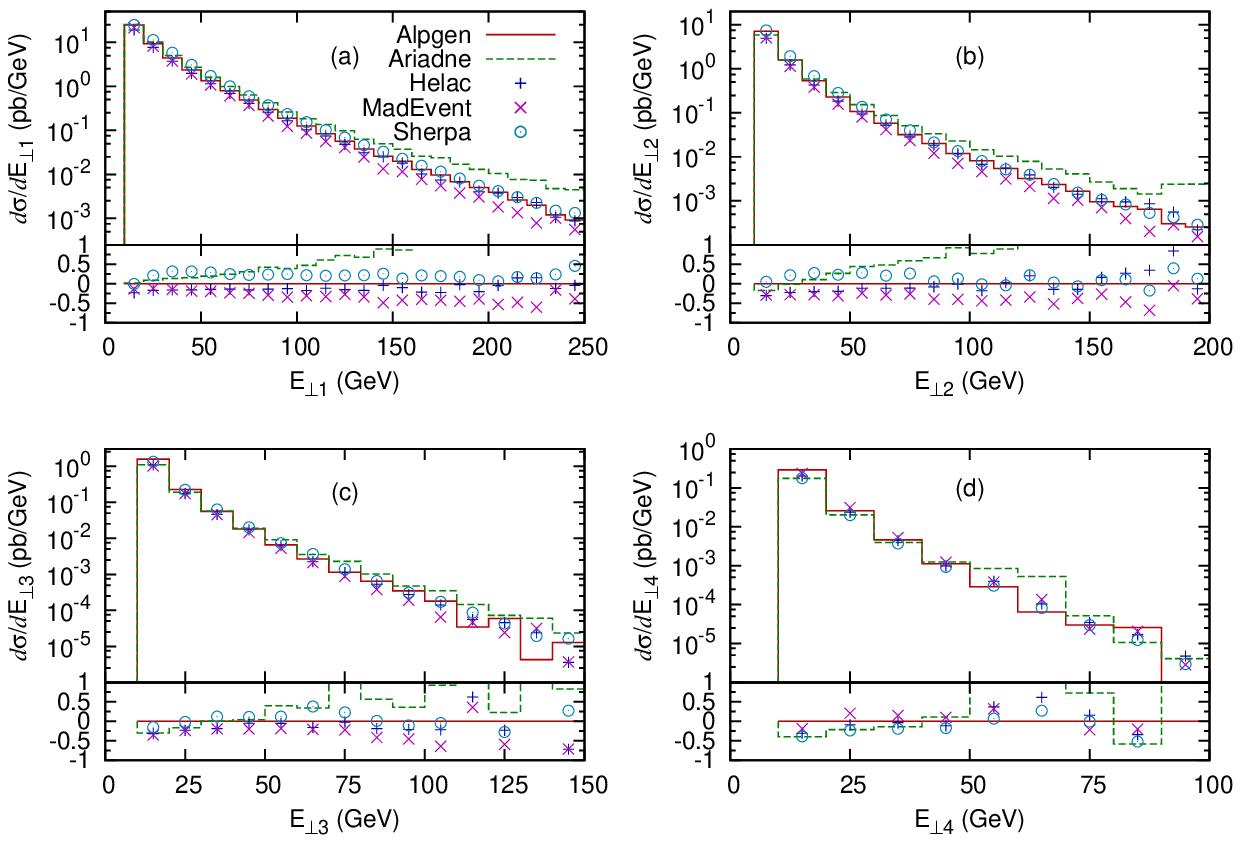}
  \end{center}
  \vskip -0.4cm \ccaption{}{\label{fig:pt-tev}Inclusive \Eperp\ spectra
    of the leading 4 jets at the Tevatron (pb/GeV). In all cases the
    full line gives the \protect\alpgen results, the dashed line gives
    the \protect\ariadne results and the ``+'', ``x'' and
    ``o'' points give the \helac, \madevent and \sherpa
    results, respectively.}
\end{figure}

We start by showing in \fig{fig:pt-tev} the inclusive \Eperp\
spectra of the leading 4 jets.  The absolute rate predicted by each
code is used, in units of pb/GeV.  The relative differences with
respect to the \alpgen results, in this figure and all other figures
of this section, are shown in the lower in-sets of each plot, where
for the code $X$\/ we plot the quantity $(\sigma(X)-\sigma_0)/\sigma_0$,
$\sigma_0$ being the values of the \alpgen curves.

There is generally good agreement between the codes, except for
\ariadne, which has a harder \Eperp\ spectra for the leading two jets.
There we also find that \sherpa is slightly harder than \alpgen and
\helac, while \madevent is slightly softer.

\begin{figure}[t!]
\begin{center}
\includegraphics[width=0.92\textwidth,clip]{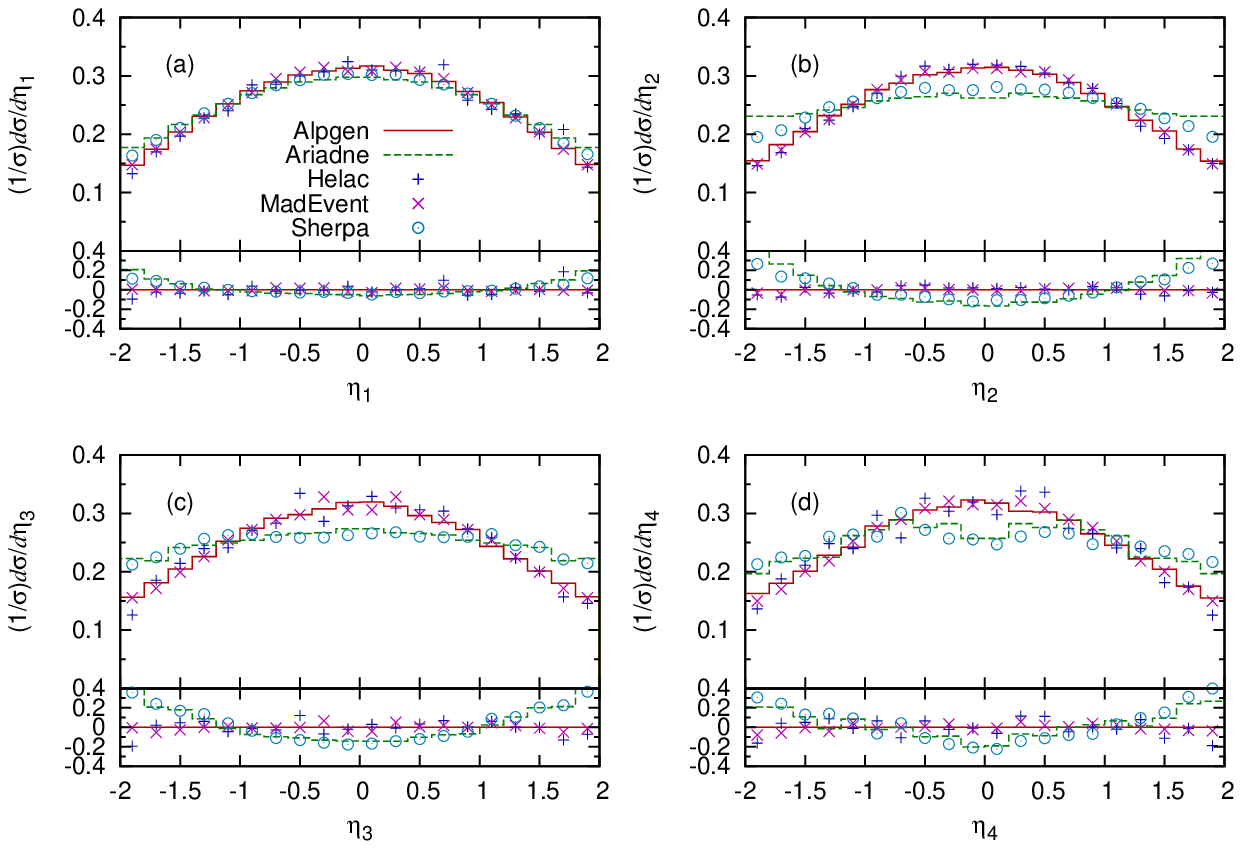}
\end{center}
\vskip -0.4cm \ccaption{}{\label{fig:eta-tev}Inclusive $\eta$\/ spectra
  of the 4 leading jets at the Tevatron. All curves are normalized to
  unit area. Lines and points are as in \fig{fig:pt-tev}.}
\vskip 0.8cm
\end{figure}

\Fig{fig:eta-tev} shows the inclusive $\eta$\/ spectra of the leading 4
jets, all normalized to unit area. There is a good agreement between
the spectra of \alpgen, \helac and \madevent, while \ariadne and
\sherpa spectra appear to be broader, in particular for the
sub-leading jets. This broadening is expected for \ariadne since the
gluon emissions there are essentially unordered in rapidity, which
means that the Sudakov form factors applied to the
matrix-element-generated states include also a $\log 1/x$\/
resummation absent in the other programs.

\begin{figure}[t!]
\begin{center}
\includegraphics[width=0.92\textwidth,clip]{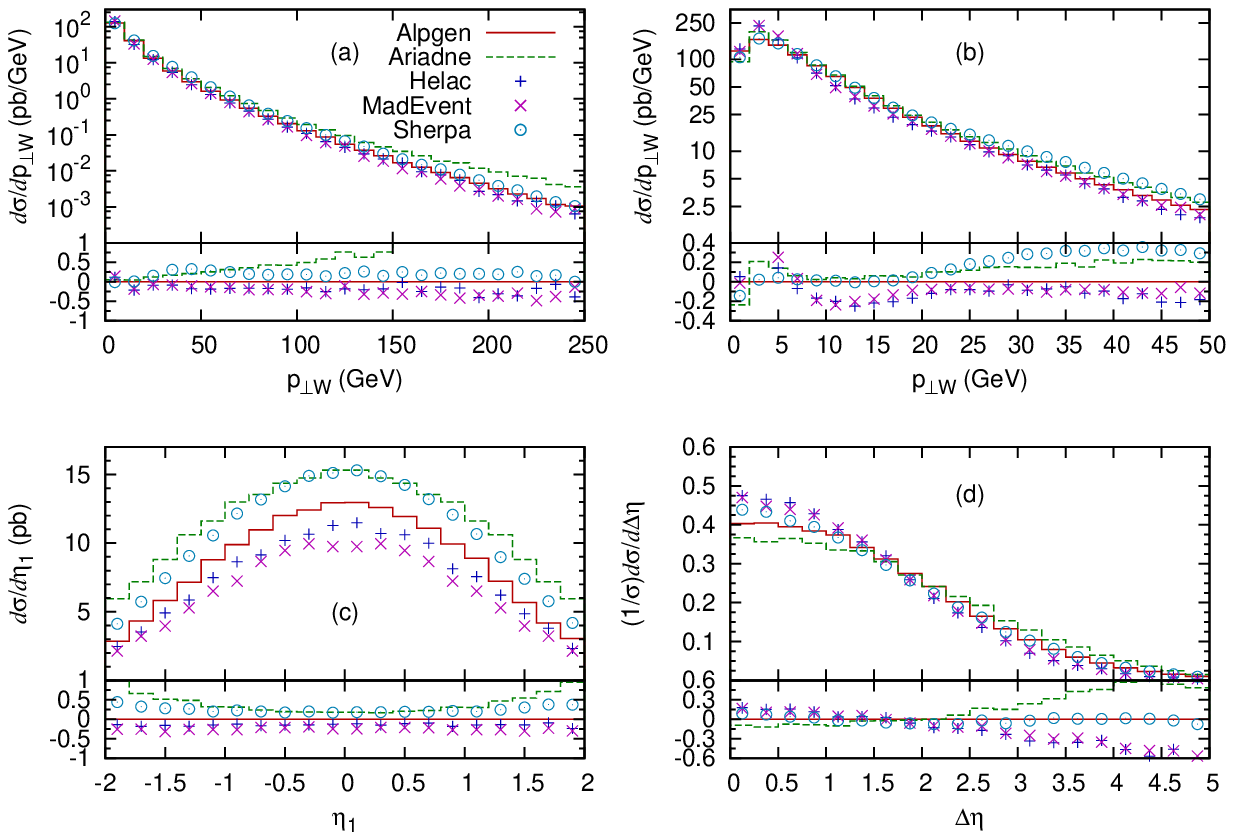}
\end{center}
\vskip -0.4cm \ccaption{}{\label{fig:ptw-tev}(a) and (b) \pperp\ spectrum
  of $W^\pm$ bosons at the Tevatron (pb/GeV). (c) Inclusive $\eta$\/
  spectrum of the leading jet, for \ppj$>50$~GeV;
  absolute normalization (pb).  (d) Pseudo-rapidity separation between
  the $W$\/ and the leading jet, $\Delta\eta=|\eta_W-\eta_{{\rm
  jet}1}|$, for \ppj$>30$~GeV, normalized to unit area.  Lines and
  points are as in \fig{fig:pt-tev}.}
\vskip 0.8cm
\end{figure}

\Fig{fig:ptw-tev}a shows the inclusive \pperp\ distribution of the $W$\/
boson, with absolute normalization in pb/GeV.  This distribution
reflects in part the behaviour observed for the spectrum of the
leading jet, with \ariadne harder than \sherpa, which, in
turn, is slightly harder than \alpgen, \helac and \madevent.  The
region of low momenta, $\ppw<50$~GeV, is expanded in
\fig{fig:ptw-tev}b. \Fig{fig:ptw-tev}c shows the $\eta$\/ distribution
of the leading jet, $\eta_1$, when its transverse momentum is larger than
50~GeV. The curves are absolutely normalized, so that it is clear how
much rate is predicted by each code to survive this harder jet cut.
The $\vert\eta\vert$ separation between the $W$\/ and the leading jet
of the event above 30 GeV is shown in \fig{fig:ptw-tev}d,
normalized to unit area.  Here we find that \ariadne has a broader
correlation, while \helac and \madevent are somewhat more narrow than
\alpgen and \sherpa.

\begin{figure}
\begin{center}
\includegraphics[width=0.92\textwidth,clip]{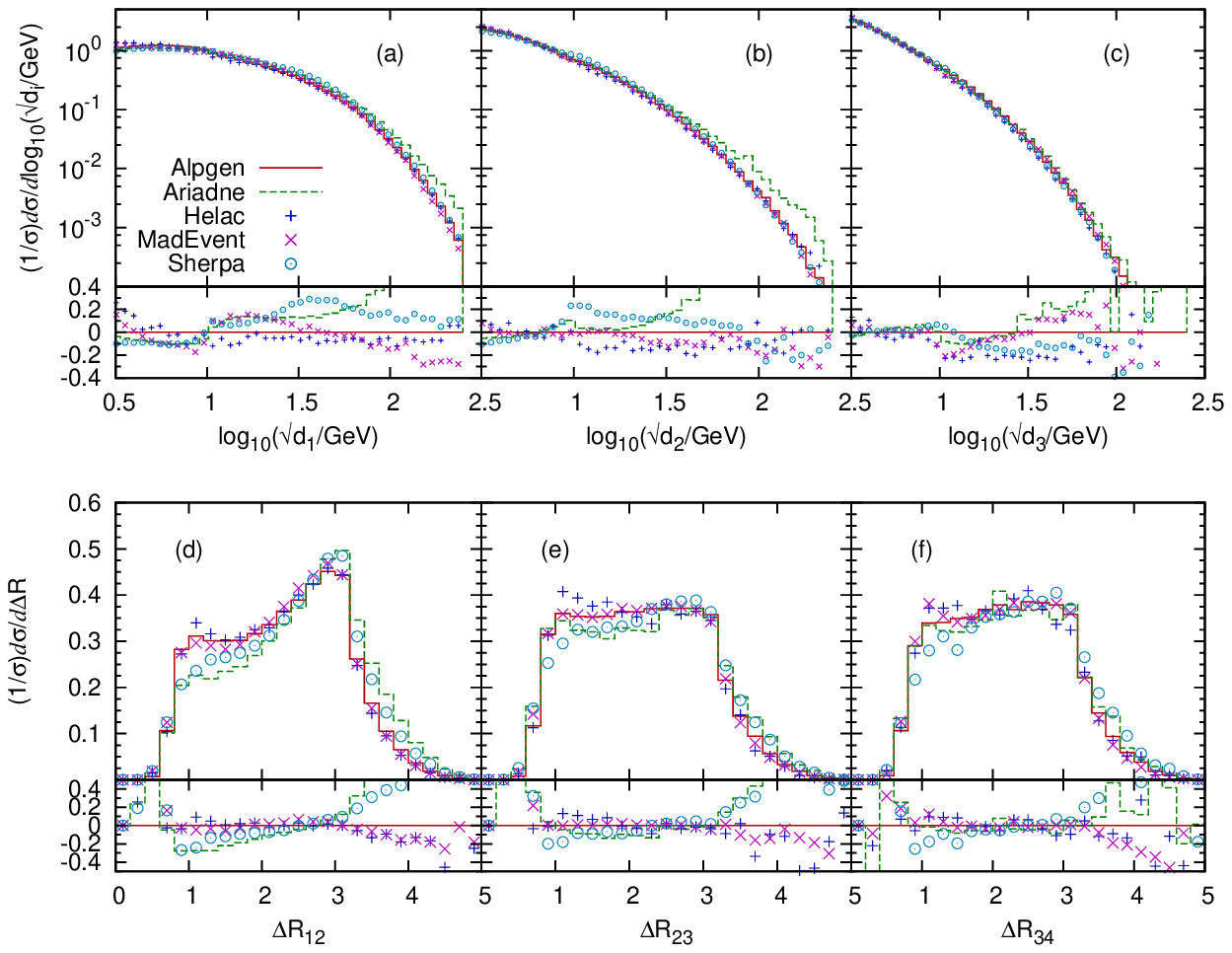}
\end{center}
\vskip -0.4cm \ccaption{}{\label{fig:dr-tev} (a)--(c) $d_i$
  ($i=1,2,3$) spectra, where $d_i$ is the scale in a parton-level
  event where $i$\/ jets are clustered into $i-1$ jets using the
  \kperp-algorithm.  (d)--(f) $\Delta R$\/ separations at the Tevatron
  between jet 1 and 2, 2 and 3, and 3 and 4. All curves are
  normalized to unit area. Lines and points are as in
  \fig{fig:pt-tev}.}
\end{figure}

\begin{figure}
\begin{center}
\includegraphics[width=0.60\textwidth,clip]{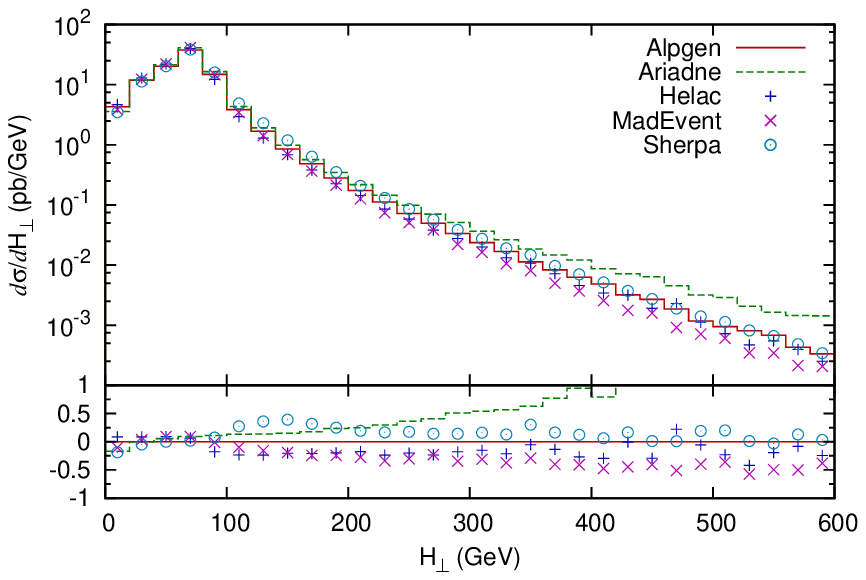}
\end{center}
\vskip -0.4cm \ccaption{}{\label{fig:ht-tev}
    The scalar sum of the transverse momentum of the charged lepton, the
    neutrino and the jets at the Tevatron. Lines and points are as in
    \fig{fig:pt-tev}.}
\end{figure}

In \fig{fig:dr-tev} we show the merging scales $d_i$ as obtained from
the \kperp-algorithm, where $d_i$ is the scale in an event where $i$\/
jets are clustered into $i-1$ jets. These are parton-level
distributions and are especially sensitive to the behaviour of the
merging procedure close to the merging/matching scale.  Note that in
the plots showing the difference the wiggles stem from both the
individual codes and from the \alpgen reference. In
section~\ref{sec:syst} below, the behaviour of the individual codes is
treated separately.

Also shown in \fig{fig:dr-tev} is the separation in $\Delta
R=\sqrt{\Delta\eta^2+\Delta\phi^2}$ between successive jet pairs
ordered in hardness. The $\Delta R_{12}$ is dominated by the
transversal-plane back-to-back peak at $\Delta R_{12}=\pi$,
while for larger $\Delta R$\/ in all cases the behaviour is more
dictated by the correlations in pseudo-rapidity. For these larger
values we find a weaker correlation in \ariadne and \sherpa, which can
be expected from their broader rapidity distributions in
\fig{fig:eta-tev}.

Finally, in \fig{fig:ht-tev} we show $H_\perp$, the scalar sum of the
transverse momenta of the charged lepton, the neutrino and the jets.
This is a variable in which one often does experimental cuts in
searches for new phenomena and is not expected to be very sensitive to
the particulars in the merging schemes. The results show good
agreement below 100 GeV, but at higher values, as expected from the
differences in the hardness of the jet and $\ppw$ spectra, \ariadne
has a harder spectra than \sherpa and \alpgen, while \madevent and
\helac has a slightly softer spectra.


\section{LHC Studies}
\label{sec:lhc-results}
\subsection{Event rates}
\label{sec:lhcrates}
\begin{figure}
\begin{center}
\includegraphics[width=0.8\textwidth,clip]{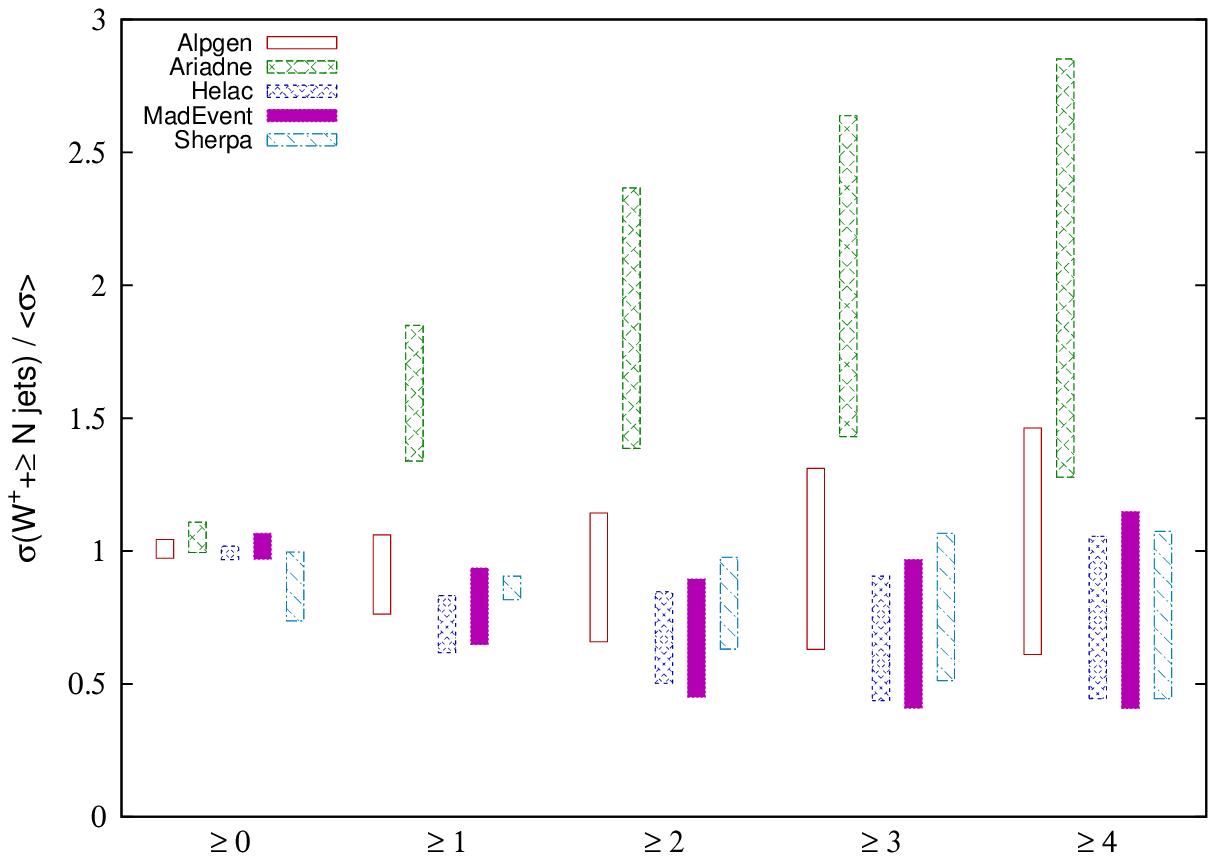}
\end{center}
\vskip -0.4cm
\ccaption{}{\label{fig:rates-lhc}Range of variation for the LHC
  cross-section rates of the five codes, normalized to the average
  value of the default settings for all codes in each multiplicity
  bin.}
\end{figure}

The tables (table~\ref{tab:lhcrates} and \ref{tab:lhcratios}) and figure
(\fig{fig:rates-lhc}) of this section parallel those shown earlier for
the Tevatron. The largest rate variations is, similarly to the Tevatron
rates, determined by the scale changes (described in more detail in
section~\ref{sec:syst}). The main feature of the LHC results is the
significantly larger rates predicted by \ariadne (see also the
discussion of its systematics, section~\ref{sec:ari-syst}), which are
outside the systematics ranges of the other codes. Aside from this and
the fact that \sherpa gives a smaller total cross section (see also the
last part of the discussion of the \sherpa systematics in
section~\ref{sec:she-syst}), the comparison among the other codes shows
an excellent consistency, with a pattern of the details similar to what
seen for the Tevatron.

\begin{table}
\begin{center}
\begin{tabular}{|l|rrrrr|}
\hline
Code & $\sigma$[tot] & $\sigma$[$\ge 1$ jet] & $\sigma$[$\ge 2$ jet]
&$\sigma$[$\ge 3$ jet] &$\sigma$[$\ge 4$ jet]  \\
\hline
    {\bf \alpgen, def } & {\bf 10170} & {\bf 2100} & {\bf 590} & {\bf
    171} & {\bf 50  }\\
ALpt30        & 10290 & 2200 & 555 & 155 & 46 \\
ALpt40        & 10280 & 2190 & 513 & 136 & 41 \\
ALscL        & 10590 & 2520 & 790 & 252 & 79 \\
ALscH        &  9870 & 1810 & 455 & 121 & 33 \\
\hline
{\bf \ariadne, def} & {\bf 10890} & {\bf 3840} & {\bf 1330} & {\bf
  384} & {\bf 101}  \\
ARpt30 & 10340 & 3400 & 1124 & 327 & 88\\
ARpt40 & 10090 & 3180 & 958 & 292 & 83 \\
ARscL  & 11250 & 4390 & 1635 & 507 & 154\\
ARscH  & 10620 & 3380 & 1071 & 275 & 69\\
ARs & 11200 & 3440 & 1398 & 438 & 130 \\

\hline {\bf \helac, def}
& {\bf 10050} & {\bf  1680} & {\bf 442} & {\bf 118} &{\bf 36} \\
HELpt40 & 10150 & 1760 & 412 & 116 & 37 \\
HELscL & 10340 & 1980 & 585 & 174 & 57 \\
HELscH &  9820 & 1470  & 347   & 84  & 24 \\
\hline
{\bf \madevent, def}  & {\bf 10830} & {\bf 2120} & {\bf 519} & {\bf
  137} & {\bf 42}  \\
MEkt30           & 10080 & 1750 & 402 & 111 & 37 \\
MEkt40           & 9840 & 1540 & 311 & 78.6 & 22 \\
MEscL           & 10130 & 2220 & 618 & 186 & 62 \\
MEscH           & 10300 & 1760 & 384 & 91.8 & 27 \\
\hline
{\bf \sherpa, def} & 
{\bf 8800} & {\bf 2130}  & {\bf 574} &  {\bf 151} &   {\bf 41} \\
SHkt30   & 8970 & 2020  & 481 &  120 &  32 \\
SHkt40   & 9200 & 1940  & 436 &   98.5 & 24 \\
SHscL    & 7480 & 2150  & 675 &  205  & 58 \\
SHscH    & 10110 & 2080 &  489 &  118 &  30 \\

\hline
\end{tabular}
\ccaption{}{\label{tab:lhcrates} Cross sections (in pb) for the
  inclusive jet rates at the LHC, according to the default and
  alternative settings of the various codes.}
\end{center}
\vskip 0.8cm
\end{table}

\begin{table}
\begin{center}
\begin{tabular}{|l|rrrr|}
\hline
Code & $\sigma^{[\ge 1]}$/$\sigma^{[\mathrm tot]}$ &
       $\sigma^{[\ge 2]}$/$\sigma^{[\ge 1]}$ &
       $\sigma^{[\ge 3]}$/$\sigma^{[\ge 2]}$ &
       $\sigma^{[\ge 4]}$/$\sigma^{[\ge 3]}$ \\
\hline
{\bf \alpgen, def} & {\bf 0.21} & {\bf 0.28} &
{\bf 0.29} & {\bf 0.29} \\
ALpt30        & 0.21 & 0.25 & 0.28 & 0.30 \\
ALpt40        & 0.21 & 0.23 & 0.27 & 0.30 \\
ALscL        & 0.24 & 0.31 & 0.32 & 0.31 \\
ALscH        & 0.18 & 0.25 & 0.27 & 0.27 \\
\hline
{\bf \ariadne, def} & {\bf 0.35} & {\bf 0.35} &
{\bf 0.29} & {\bf 0.26} \\
ARpt30           & 0.33 & 0.33 & 0.29 & 0.27 \\
ARpt40           & 0.32 & 0.30 & 0.30 & 0.28 \\
ARscL             & 0.39 & 0.37 & 0.31 & 0.30 \\
ARscH             & 0.32 & 0.32 & 0.26 & 0.24 \\
ARs           & 0.31 & 0.41 & 0.31 & 0.30 \\
\hline
{\bf \helac, def} &  {\bf 0.17} &  {\bf 0.26} &
{\bf 0.27} &  {\bf 0.31} \\
HELpt40 &  0.17 &  0.23 &  0.28 &  0.32 \\
HELscL &  0.19 &  0.30 &  0.30 &  0.33 \\
HELscH &  0.15 &  0.24 &  0.24 &  0.29 \\
\hline
{\bf \madevent, def} & {\bf 0.20} & {\bf 0.24} &
{\bf 0.26} & {\bf 0.31}\\
MEkt30          & 0.17 & 0.23 & 0.28 & 0.33 \\
MEkt40          & 0.16 & 0.20 & 0.25 & 0.28 \\
MEscL          & 0.22 & 0.27 & 0.30 & 0.34 \\
MEscH          & 0.17 & 0.22 & 0.24 & 0.29 \\
\hline
{\bf \sherpa, def} & {\bf 0.24} &  {\bf 0.27} &
{\bf 0.26} &  {\bf 0.27} \\
SHkt30          &  0.23 &  0.24 &  0.25 &  0.27 \\
SHkt40          &  0.21 &  0.22 &  0.23 &  0.24 \\
SHscL          &  0.29 &  0.31 &  0.30 &  0.28 \\
SHscH          &  0.21 &  0.24 &  0.24 &  0.25 \\
\hline
\end{tabular}
\ccaption{}{\label{tab:lhcratios} Cross-section ratios for $(n+1)/n$\/
  inclusive jet rates at the LHC, according to the default and
  alternative settings of the various codes.}
\end{center}
\end{table}

\subsection{Kinematical distributions}
\label{sec:lhcdists}

\begin{figure}
\begin{center}
\includegraphics[width=0.9\textwidth,clip]{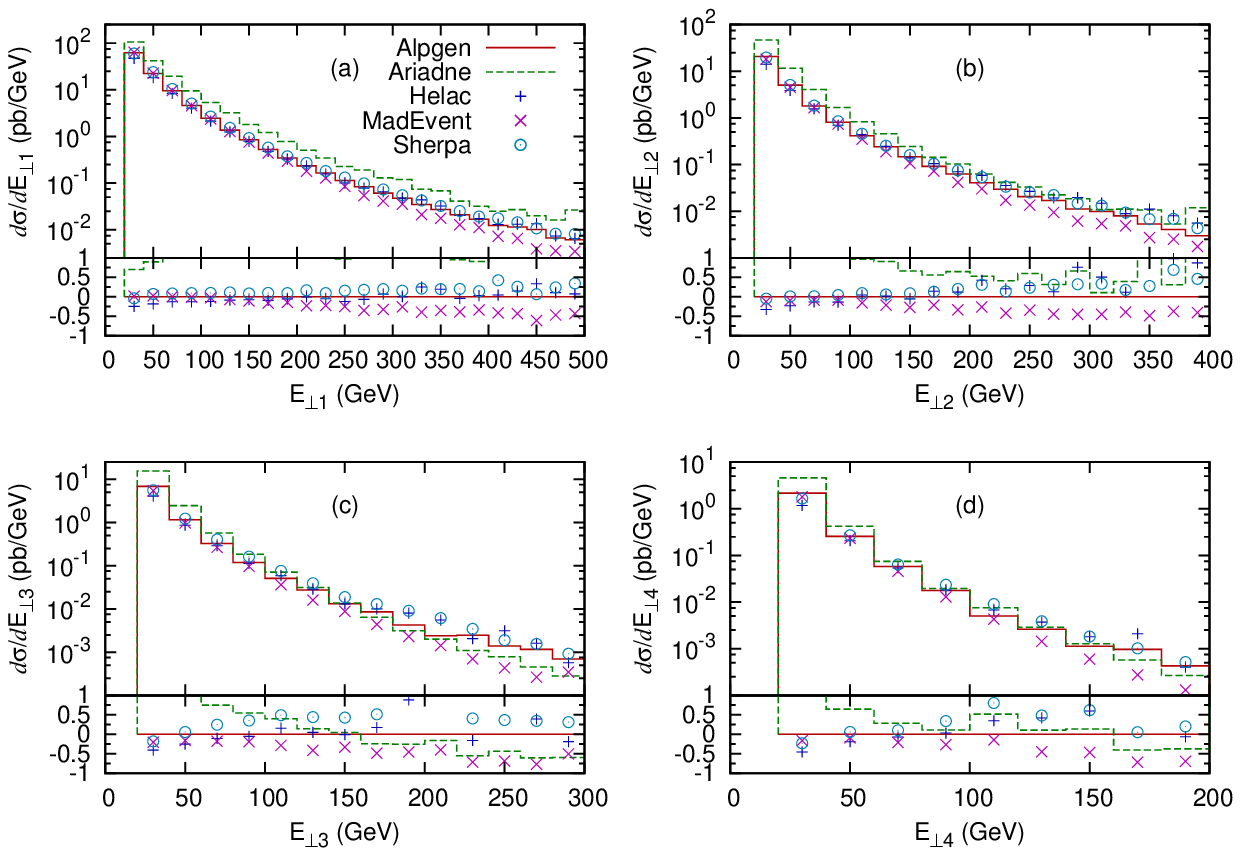}
\end{center}
\vskip -0.9cm \ccaption{}{\label{fig:pt-lhc}Inclusive \Eperp\ spectra
  of the leading 4 jets at the LHC (pb/GeV). In all cases the full
  line gives the \protect\alpgen results, the dashed line gives the
  \protect\ariadne results and the ``+'', ``x'' and ``o'' points give
  the \helac, \madevent and \sherpa results respectively.}
\end{figure}

Following the same sequence of the Tevatron study, we start by showing
in \fig{fig:pt-lhc} the inclusive \Eperp\ spectra of the leading 4
jets.  The absolute rate predicted by each code is used, in units of
pb/GeV.

Except for \ariadne, we find good agreement among the codes, with
\ariadne having significantly harder leading jets, while for
sub-leading jets the increased rates noted in
\fig{fig:rates-lhc} mainly come from lower \Eperp. Among the
other codes, \helac and \sherpa have consistently somewhat harder jets
than \alpgen, while \madevent is a bit softer, but these differences
are not as pronounced.

\begin{figure}
\vskip -0.4cm
\begin{center}
\includegraphics[width=0.9\textwidth,clip]{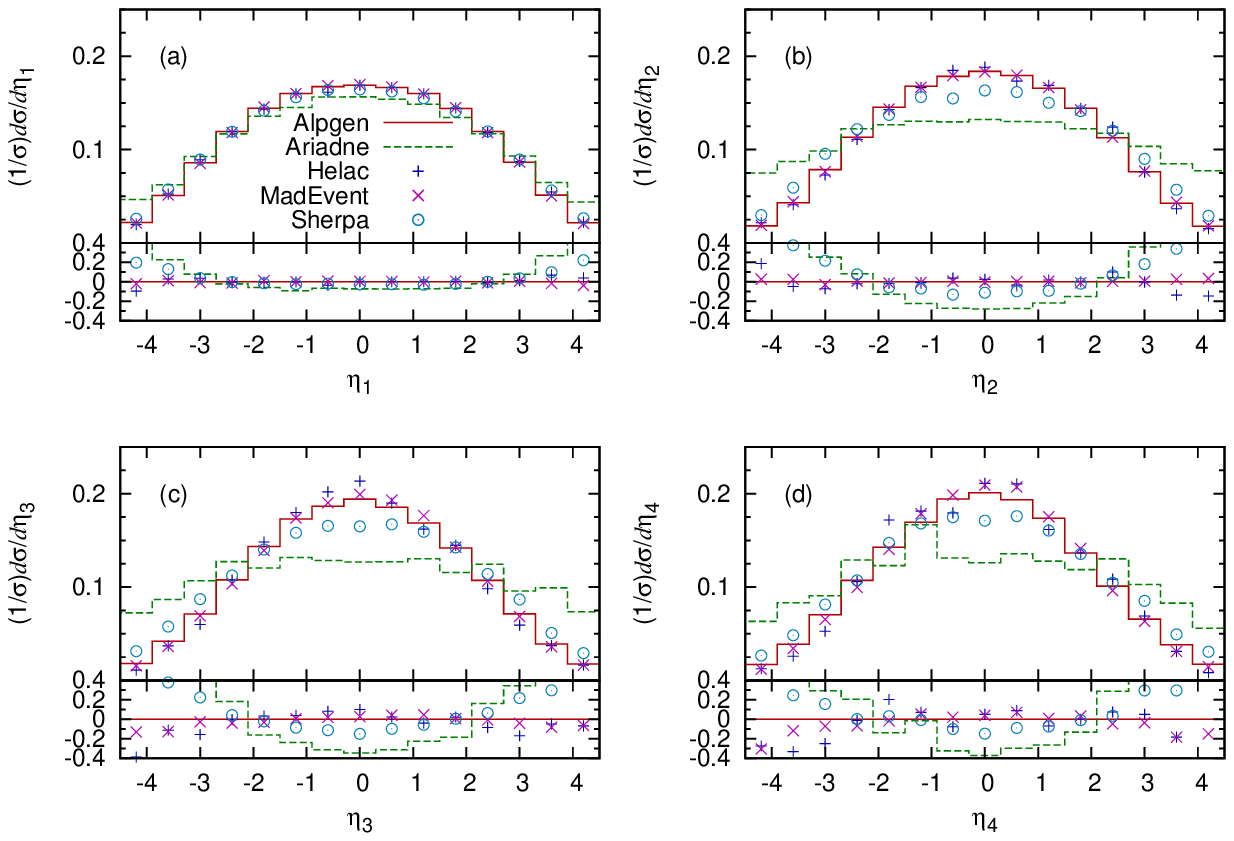}
\end{center}
\vskip -0.9cm
\ccaption{}{\label{fig:eta-lhc}Inclusive $\eta$\/ spectra of the 4 leading
  jets at the LHC. All curves are normalized to unit area. Lines and
  points are as in \fig{fig:pt-lhc}.}
\end{figure}
\begin{figure}
\begin{center}
\includegraphics[width=0.92\textwidth,clip]{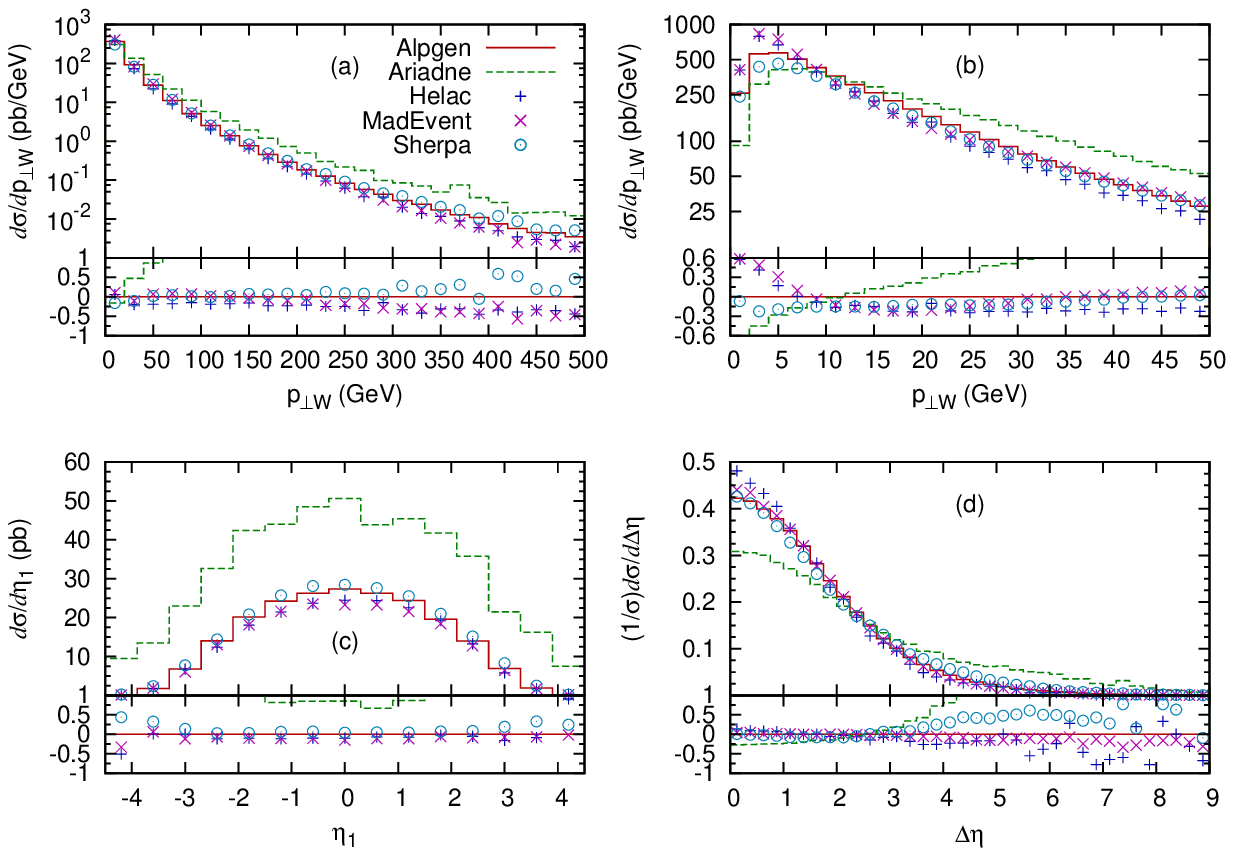}
\end{center}
\vskip -0.9cm \ccaption{}{\label{fig:ptw-lhc} (a) and (b) \pperp\
  spectrum of $W^+$ bosons at the LHC (pb/GeV). (c) $\eta$\/ spectrum
  of the leading jet, for \ppj$>100$~GeV; absolute normalization (pb).
  (d) Pseudo-rapidity separation between the $W^+$ and the leading
  jet, $\Delta\eta=|\eta_{W^+}-\eta_{{\rm jet}1}|$, for
  \ppj$>40$~GeV, normalized to unit area. Lines and points are as in
  \fig{fig:pt-lhc}.}
\vskip -0.3cm
\end{figure}

For the pseudo-rapidity spectra of the jets in \fig{fig:eta-lhc} it is
clear that \ariadne has a much broader distribution in all cases. Also
\sherpa has broader distributions, although not as pronounced, while
the other codes are very consistent.

The \pperp\ distribution of $W^+$ bosons in \fig{fig:ptw-lhc} follows
the trend of the leading-jet \Eperp\ spectra. The differences observed
in the \ppw\ region below 10 GeV are not due to the choice of merging
approach, but are entirely driven by the choice of shower algorithm.
Notice for example the similarity of the \helac and \madevent spectra,
and their peaking at lower pt than the \herwig spectrum built into the
\alpgen curve, a result well known from the comparison of the standard
\pythia and \herwig generators. Increasing the transverse momentum of
the leading jet in \fig{fig:ptw-lhc}a does not change the conclusions
much for its pseudo-rapidity distribution. Also the rapidity correlation
between the leading jet and the $W^+$ follows the trend found for the
Tevatron, but the differences are larger, with a much weaker correlation
for \ariadne. Also \sherpa shows a somewhat weaker correlation, while
\helac is somewhat stronger than \alpgen and \madevent.

\begin{figure}
\begin{center}
\includegraphics[width=0.92\textwidth,clip]{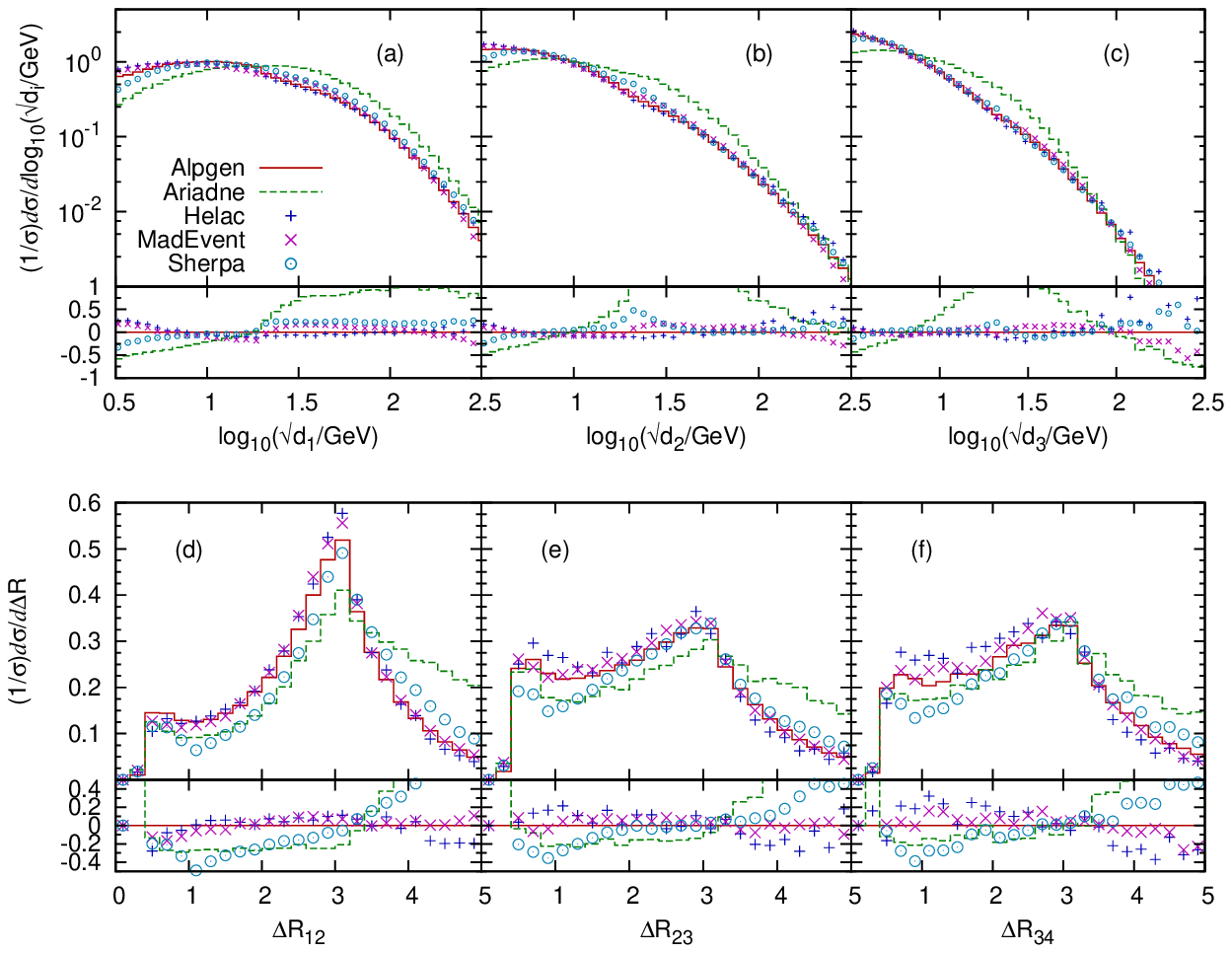}
\end{center}
\vskip -0.4cm \ccaption{}{\label{fig:dr-lhc} (a)--(c) $d_i$
  ($i=1,2,3$) spectra, where $d_i$ is the scale in a parton-level
  event where $i$ jets are clustered into $i-1$ jets using the
  \kperp-algorithm.  (d)--(f) $\Delta R$ separations at the LHC
  between jet 1 and 2, 2 and 3, and 3 and 4. All curves are
  normalized to unit area. Lines and points are as in
  \fig{fig:pt-lhc}.}
\end{figure}

\begin{figure}
\begin{center}
\includegraphics[width=0.60\textwidth,clip]{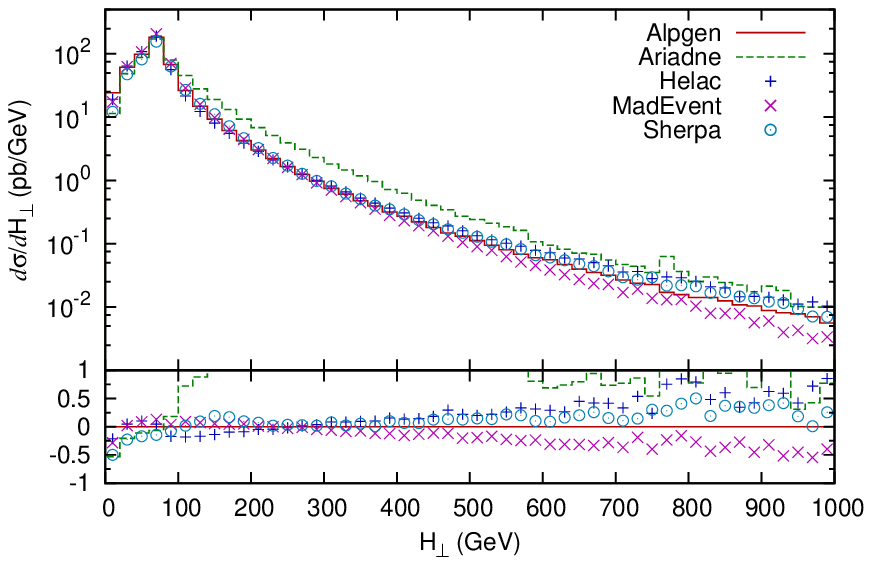}
\end{center}
\vskip -0.4cm \ccaption{}{\label{fig:ht-lhc}
    The scalar sum of the transverse momentum of the charged lepton, the
    neutrino and the jets at the LHC. Lines and points are as in
    \fig{fig:pt-lhc}.}
\end{figure}

For the distribution in clustering scale in \fig{fig:dr-lhc}, we find
again that \ariadne is by far the hardest. The results given by the
other codes are comparable, with the only exception that for the $d_1$
distribution, \sherpa gives a somewhat harder prediction compared to
the ones made by the MLM-based approaches.

The $\Delta R$ distributions, in \fig{fig:dr-lhc}, show at large
separation a behaviour consistent with the broad rapidity distributions
found for \sherpa, and in particular for \ariadne, in \fig{fig:eta-lhc}.
This increase at large $\Delta R$ is then compensated by a depletion
with respect to the other codes at small separation.

The scalar transverse momentum sum in \fig{fig:ht-lhc} shows
significantly larger deviations as compared to the results for the
Tevatron. \ariadne has a much harder spectra than the other codes,
while \sherpa and \helac are slightly harder than \alpgen and
\madevent is significantly softer. As in the Tevatron case, it is a
direct reflection of the differences in the hardness of the jet and
$\ppw$ spectra, although the increased phase space for jet production
at the LHC makes the $\ppw$ contribution less important at high
$H_\perp$ values.


\section{Systematic studies}
\label{sec:syst}

In this section we present the systematic studies of each of the codes
separately for both the Tevatron and the LHC, followed by some general
comments on differences and similarities between the codes.

In all cases we have chosen a subset of the plots shown in the
previous sections: the transverse momentum of the $W$, the
pseudo-rapidity of the leading jet, the separation between the leading
and the sub-leading jet, and the $d_i$ logarithmic spectra. As before,
all spectra aside from \ppw\ are normalized to unit integral over the
displayed range. The variations of the inclusive jet cross sections
has already been shown in table~\ref{tab:tevrates}-\ref{tab:lhcratios}
and \figs{fig:rates-tev} and \ref{fig:rates-lhc}.

To estimate what systematic error can be expected from each code, the
effects of varying the merging scale and changing the scale used in the
determination of the strong coupling is studied (the details for each
code is described in section~\ref{sec:genprop}). The merging scale
variations are introduced according to the definition in each algorithms
and should lead to small changes in the results, although the nonleading
terms from the matrix elements always lead to some residual dependence
on the merging scale. In the various algorithms different choices have
been made regarding how to estimate the uncertainty from \alps-scale
variations and this leads to slightly different physical consequences.

In the case of \alpgen and \helac, the scale changes are only
implemented in the strong coupling calculated in the matrix element
reweighting, but the scale in the shower remains unchanged. This leads
to variations of the result that are proportional to the relevant
power of \alps used in the matrix element, which means that the spectra
contains small deviations below the merging scale and that the
deviations grow substantially above the merging scale.

In \ariadne, \madevent and \sherpa both the scale in the
\alps-reweighting and the scale in the \alps of the shower is changed.
In addition to this the scale used in the evaluation of the parton
densities is also changed in \sherpa (this is discussed further in
section~\ref{sec:she-syst}). Including the scale variations in \alps in
the shower changes the fraction of rejected events or the Sudakov form
factors (depending on which algorithm is used), which modifies the cross
section in the opposite direction compared to the scale changes in the
matrix element reweighting. This leads to smaller deviations in the
results above the merging scale and it is also possible to get
significant deviations in the opposite direction below the merging
scale, which is mainly visible in the \ppw\ spectra.

\begin{figure}
\begin{center}
\includegraphics[width=0.92\textwidth,clip]{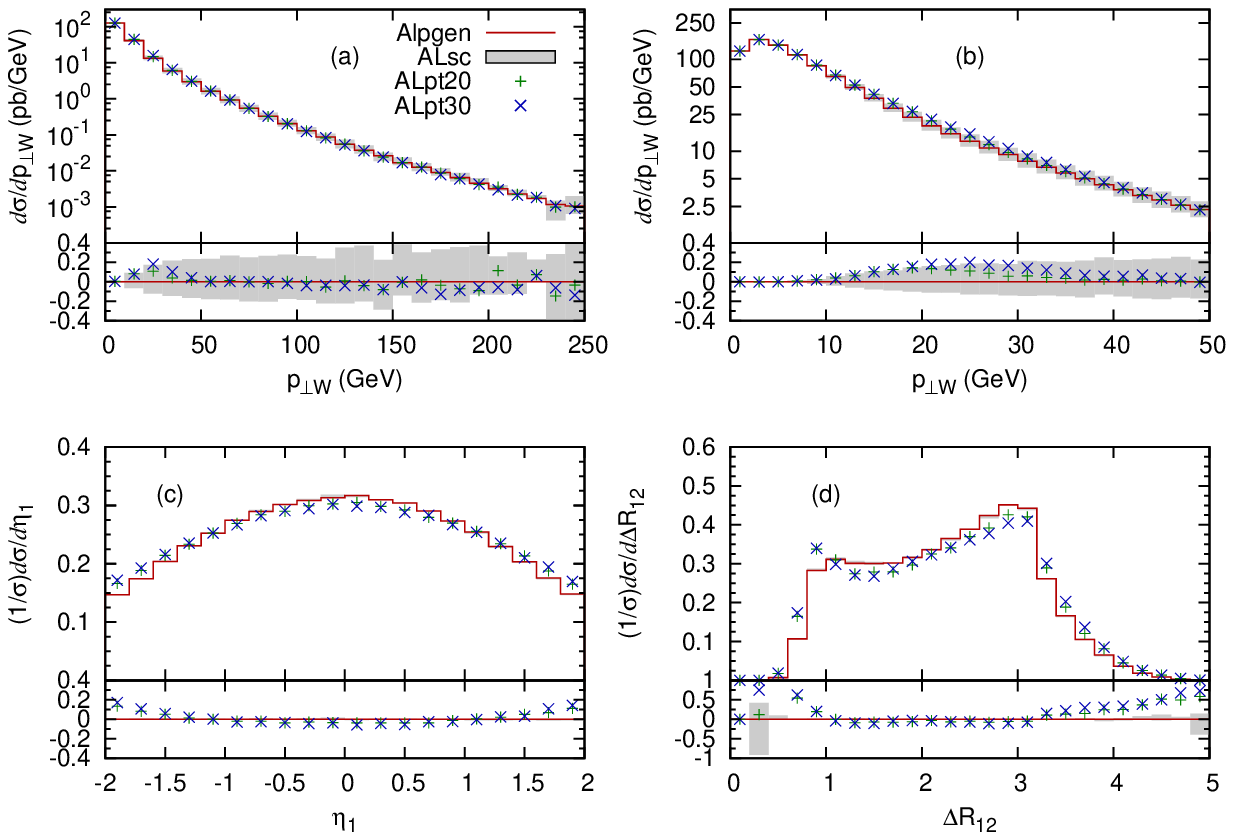}
\includegraphics[width=0.92\textwidth,clip]{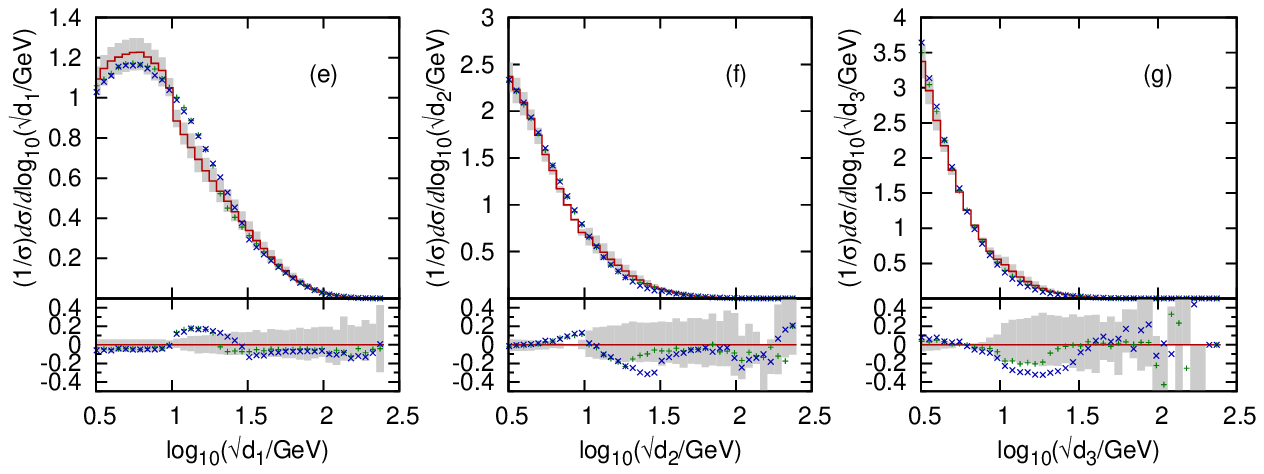}
\end{center}
\vskip -0.4cm
\ccaption{}{\label{fig:alp-ptw-tev} \protect\alpgen
  systematics at the Tevatron. (a) and (b) show the \pperp\ spectrum of
  the $W$, (c) shows the pseudo-rapidity distribution of the leading jet,
  (d) shows the $\Delta R$\/ separation between the two leading jets,
  and (e)--(g) show the $d_i$ ($i=1,2,3$) spectra, where $d_i$ is the
  scale in a parton-level event where $i$\/ jets are clustered into $i-1$
  jets using the \kperp-algorithm. The full line is the default settings
  of \protect\alpgen, the shaded area is the range between ALscL and
  ALscH, while the points represent ALpt20 and ALpt30 as defined in
  section~\ref{sec:genprop}.}
\end{figure}

\begin{figure}
\begin{center}
\includegraphics[width=0.92\textwidth,clip]{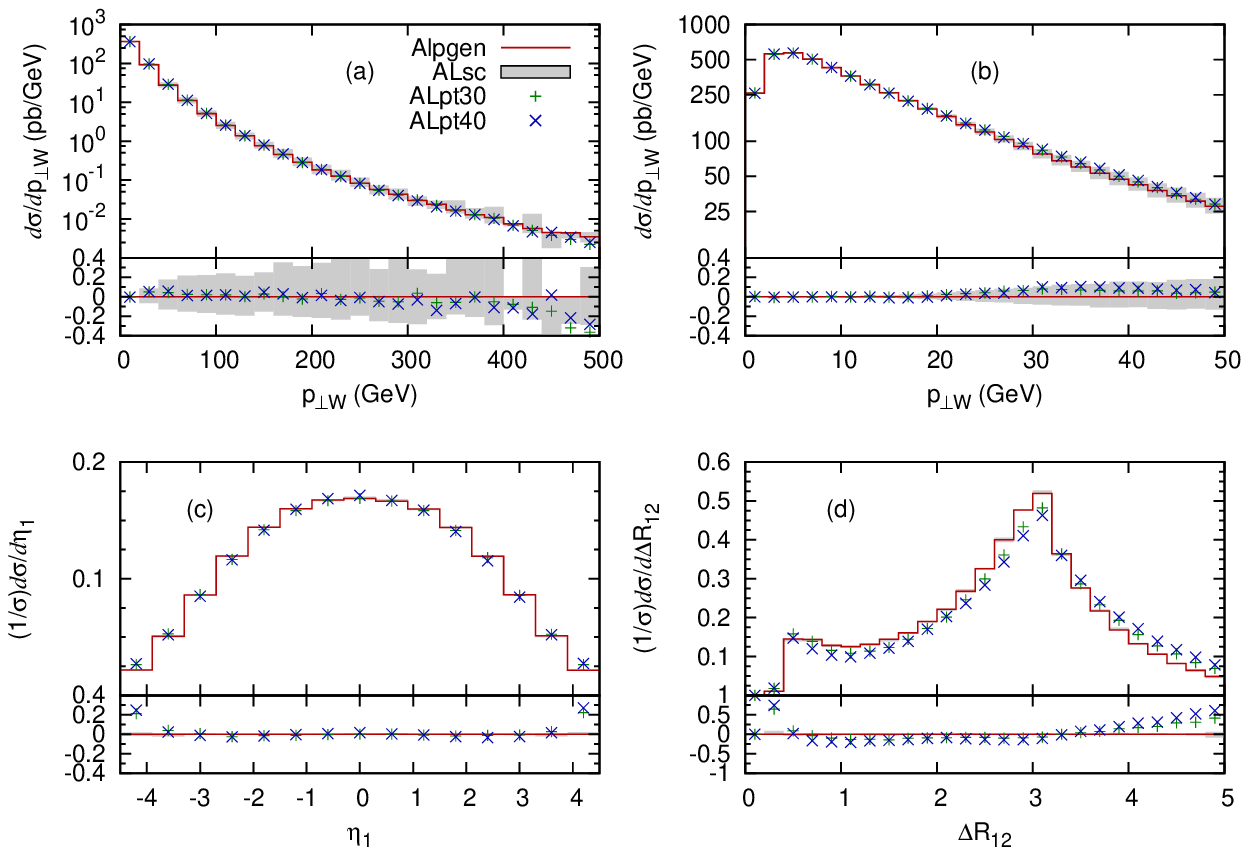}
\includegraphics[width=0.92\textwidth,clip]{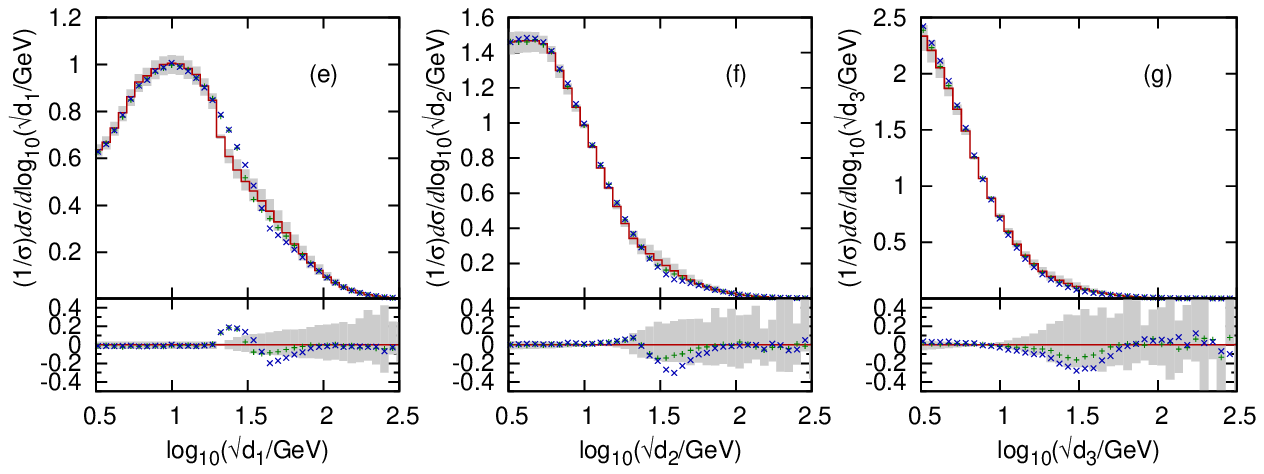}
\end{center}
\vskip -0.4cm
\ccaption{}{\label{fig:alp-ptw-lhc}\protect\alpgen systematics at the
  LHC. (a) and (b) show the \pperp\ spectrum of the $W$, (c) shows the
  pseudo-rapidity distribution of the leading jet, (d) shows the $\Delta
  R$\/ separation between the two leading jets, and (e)--(g) show the
  $d_i$ ($i=1,2,3$) spectra, where $d_i$ is the scale in a parton-level
  event where $i$ jets are clustered into $i-1$ jets using the
  \kperp-algorithm. The full line is the default settings of
  \protect\alpgen, the shaded area is the range between ALscL and ALscH,
  while the points represent ALpt30 and ALpt40 as defined in
  section~\ref{sec:genprop}.}
\end{figure}

\subsection{\protect\alpgen systematics}
\label{sec:alp-syst}

The \alpgen distributions for the Tevatron are shown in
\fig{fig:alp-ptw-tev}.  The pattern of variations is consistent
with the expectations. In the case of the \ppw\ spectra, which are
plotted in absolute scales, the larger variations are due to the
change of scale, with the lower scale leading to a harder spectrum.
The $\pm 20\%$ effect is consistent with the scale variation of
$\alpha_s$, which dominates the scale variation of the rate once \ppw\
is larger than the Sudakov region. The change of matching scales only
leads to a minor change in the region $0\ \gev <\ppw<40$~GeV, confirming
the stability of the merging prescription.

In the case of the rapidity spectrum, we notice that the scale change
leaves the shape of the distribution unaltered, while small changes
appear at the edges of the $\eta$\/ range. The $d_i$ distributions show
agreement among the various options when $\sqrt{d_i} < 10$~GeV. This is
due to the fact that the region $\sqrt{d_i} < 10$~GeV is dominated by
the initial-state evolution of an $n=i-1$ parton event, and both the
matching and scale sensitivities are reduced. Notice that in the \alpgen
prescription the scale for the shower evolution is kept fixed when the
renormalization scale of the matrix elements is changed, as a way of
exploring the impact of a possible mismatch between the two.

For $\sqrt{d_i}>\etclus$ the jet transverse energies are themselves
typically above $\etclus$, and the sensitivity to matching thresholds
smaller than $\etclus$ is reduced, since if the event matched at
$\etclus$, it will also match below that. Here the main source of
systematics is therefore the scale variation, associated to the hard
matrix element calculation for the $n=i$ jet multiplicity. The region
$10<\sqrt{d_i}<\etclus$ is the transition region between the dominance
of the shower and of the matrix element description of hard radiation.
The structure observed in the $d_i$ distributions in this region
reflects the fact that shower and matrix element emit radiation with a
slightly different probability. The selection of a matching threshold,
which leads to effects at the level of $\pm 20\%$ and is therefore
consistent with a LL accuracy and can be used to tune to data.

For the LHC, the \alpgen systematics is shown in \fig{fig:alp-ptw-lhc}.
The comparison of the various parameter choices is similar to what we
encountered at the Tevatron, with variations in the range of $\pm
20\%$ for the matching-scale systematics, and up to 40\% for the scale
systematics. The pattern of the glitches in the $d_i$ spectra for the
different matching thresholds is also consistent with the explanation
provided in the case of the Tevatron.

\subsection{\protect\ariadne systematics}
\label{sec:ari-syst}

\begin{figure}
\begin{center}
\includegraphics[width=0.92\textwidth,clip]{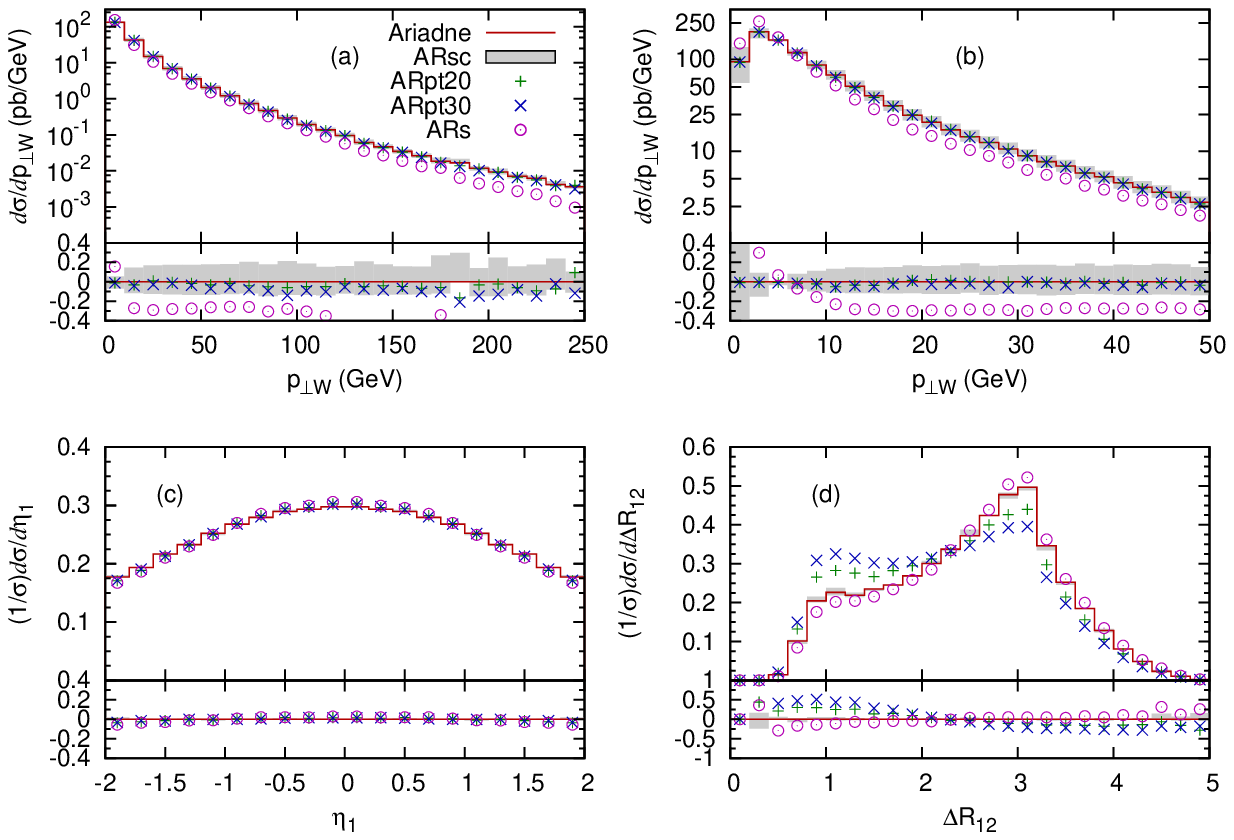}
\includegraphics[width=0.92\textwidth,clip]{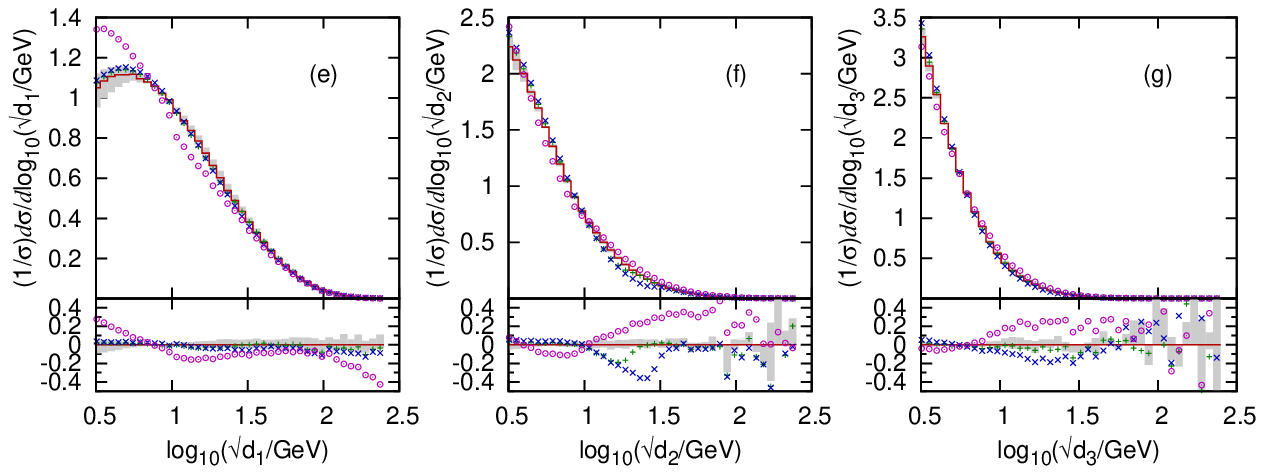}
\end{center}
\vskip -0.4cm
\ccaption{}{\label{fig:ari-ptw-tev} \protect\ariadne systematics at the
  Tevatron. The plots are the same as in \fig{fig:alp-ptw-tev}. The
  full line is the default settings of \protect\ariadne, the shaded
  area is the range between ARscL and ARscH, while the points
  represent ARpt20, ARpt30 and ARs as defined in section~\ref{sec:genprop}.}
\end{figure}

\begin{figure}
\begin{center}
\includegraphics[width=0.92\textwidth,clip]{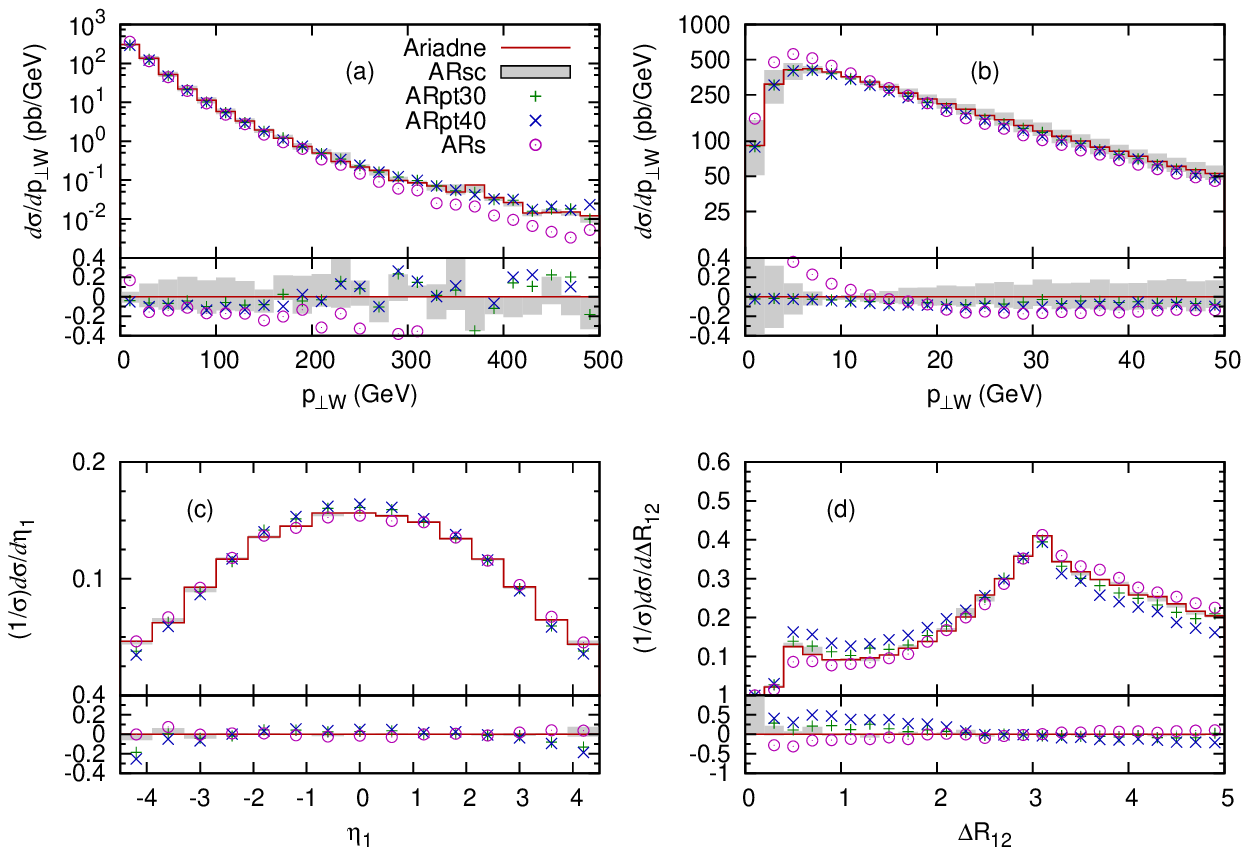}
\includegraphics[width=0.92\textwidth,clip]{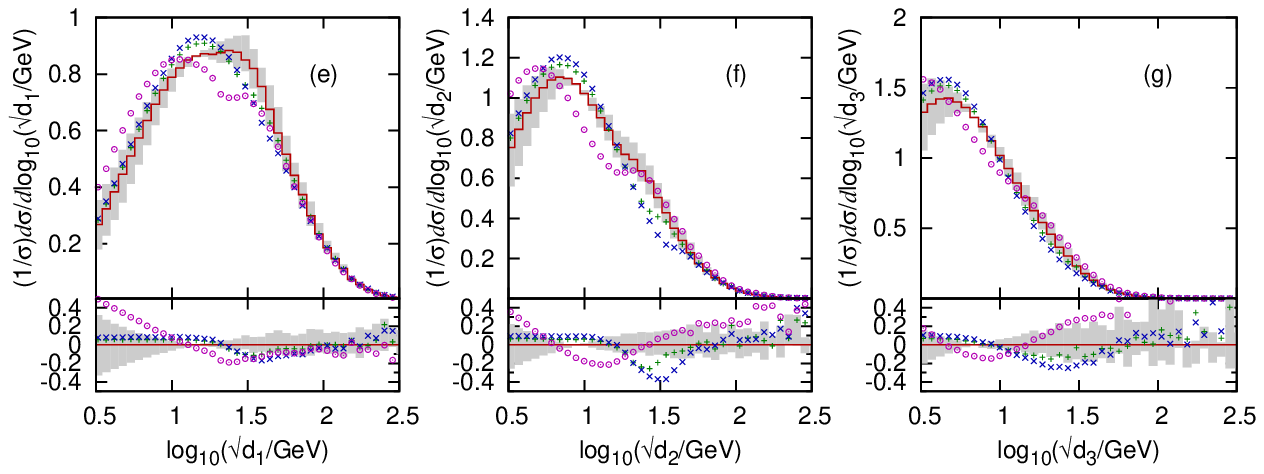}
\end{center}
\vskip -0.4cm
\ccaption{}{\label{fig:ari-ptw-lhc}\protect\ariadne systematics at the
  LHC. The plots are the same as in \fig{fig:alp-ptw-lhc}. The full
  line is the default settings of \protect\ariadne, the shaded area is
  the range between ARscL and ARscH, while the points represent
  ARpt30, ARpt40 and ARs as defined in section~\ref{sec:genprop}.}
\end{figure}

The \ariadne systematics for the Tevatron is shown in
\fig{fig:ari-ptw-tev}.  Since the dipole cascade by itself already
includes a matrix-element correction for the first emission, we see no
dependence on the merging scale in the \ppw, $\eta_{{\rm jet}1}$ and
$d_1$ distributions, which are mainly sensitive to leading order
corrections.  For the other distributions, we become sensitive to
higher-order corrections, and here the pure dipole cascade
underestimates the matrix element and also tends to make the leading
jets less back-to-back in azimuth. The first effect is expected for
all parton showers, but is somewhat enhanced in \ariadne due to the
missing initial-state $q\to gq$ splitting, and is mostly visible in
the $d_2$ distribution just below the merging scale. The second effect
is clearly visible in the $\Delta R_{12}$ distribution, which is
dominated by low \Eperp\ jets.

The changing of the soft suppression parameter in ARs has the effect
of reducing the available phase space of gluon radiation, especially
for large \Eperp\ and in the beam directions, an effect, which is mostly
visible for the hardest emission and in the \ppw\ distribution.
As for \alpgen, and also for the other codes, the change in scale
mainly affects the hardness of the jets, but not the $\eta_{{\rm
jet}1}$ and the $\Delta R_{12}$ distribution.

For the LHC, the \ariadne systematics is shown in
\fig{fig:ari-ptw-lhc}.  Qualitatively we find the same effects as in
the Tevatron case. In particular we note the strong dependence on the
soft suppression parameters in ARs, and it is clear that these have to
be adjusted to fit Tevatron (and HERA) data before any predictions for
the LHC can be made. It should be noted, however, that while eg.\ the
high \ppw\ tail in \fig{fig:ari-ptw-lhc}a for ARs is shifted down to be
comparable to the other codes (cf.\ \fig{fig:ptw-lhc}a), the medium
\ppw\ values are less affected and here the differences compared to the
other codes can be expected to remain after a retuning.

This difference is mainly due to the fact that the dipole cascade in
\ariadne, contrary to the other parton showers, is not based on
standard DGLAP evolution, but also allows for evolution, which is
unordered in transverse momentum \`a la BFKL\footnote{The dipole
  emission of gluons in \ariadne are ordered in transverse momentum,
  but not in rapidity.  Translated into a conventional initial-state
  evolution, this corresponds to emissions ordered in rapidity but
  unordered in transverse momentum.}. This means that in \ariadne
there is also a resummation of logs of $1/x$\/ besides the standard
$\log Q^2$ resummation. This should not be a large effect at the
Tevatron, and the differences there can be tuned away by changing the
soft suppression parameters in \ariadne. However, at the LHC we have
quite small $x$-values, $x\sim m_W/\sqrt{S}<0.01$, which allow for a
much increased phase space for jets as compared to what is allowed by
standard DGLAP evolution. As a result one obtains larger inclusive jet
rates as documented in table~\ref{tab:lhcrates}. The same effect is
found in DIS at HERA, where $x$\/ is even smaller as are the typical
scales, $Q^2$. And here, all DGLAP-based parton showers fail to
reproduce final-state properties, especially forward jet rates, while
\ariadne does a fairly good job.

It would be interesting to compare the merging schemes presented here
also to HERA data to see if the DGLAP based shower would better
reproduce data when merged with higher-order matrix elements. This
would also put the extrapolations to the LHC on safer grounds.
However, so far there exists one preliminary such study for the
\ariadne case only\cite{abergthesis}.

\subsection{\protect\helac systematics}
\label{sec:hel-syst}

\begin{figure}
\begin{center}
\includegraphics[width=0.92\textwidth,clip]{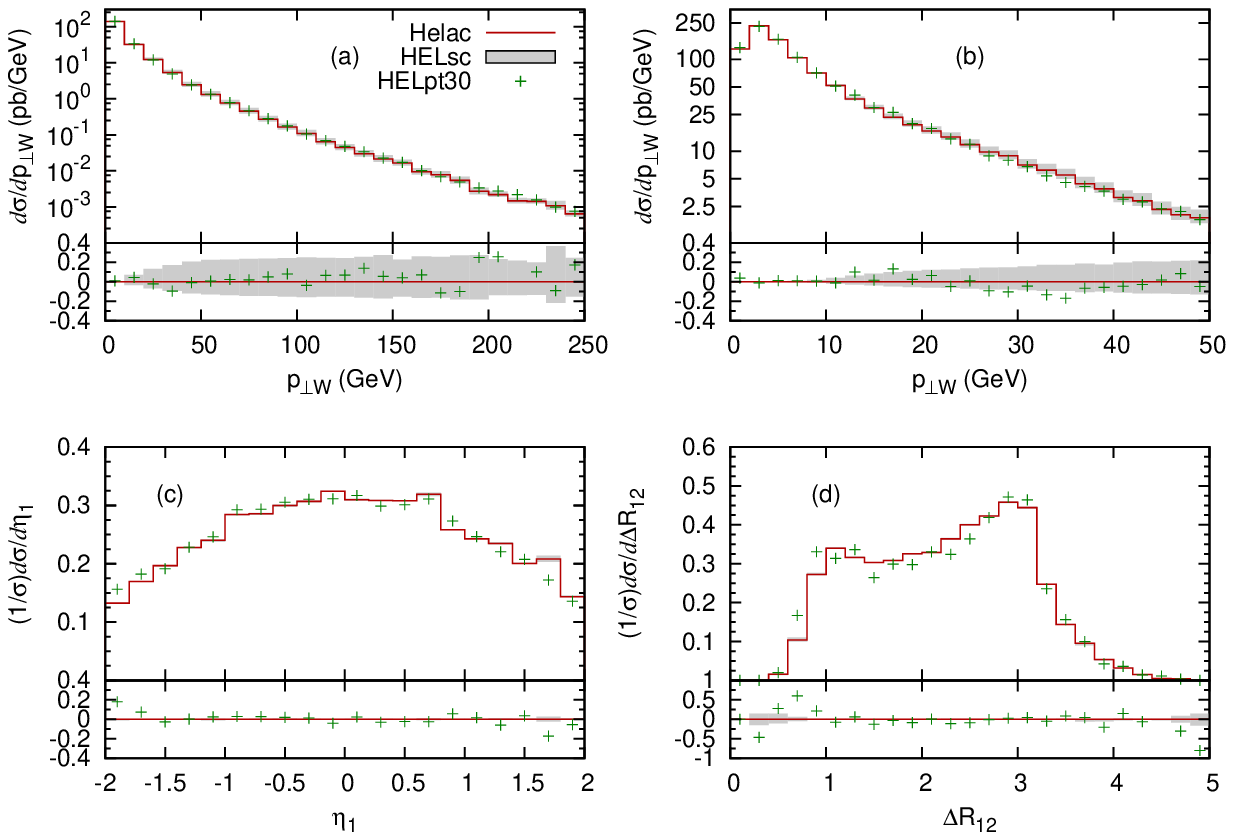}
\includegraphics[width=0.92\textwidth,clip]{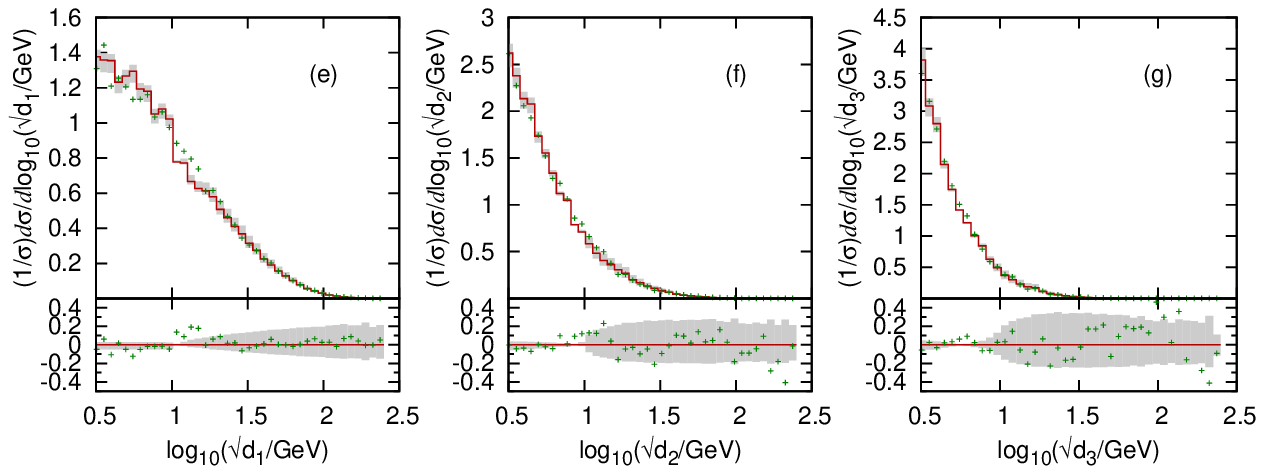}
\end{center}
\vskip -0.4cm
\ccaption{}{\label{fig:hel-ptw-tev}\protect\helac systematics at the
  Tevatron. The plots are the same as in \fig{fig:alp-ptw-tev}. The
  full line is the default settings of \protect\helac, the shaded
  area is the range between HELscL and HELscH, while the points
  represent HELpt30 as defined in section~\ref{sec:genprop}.}
\end{figure}

\begin{figure}
\begin{center}
\includegraphics[width=0.92\textwidth,clip]{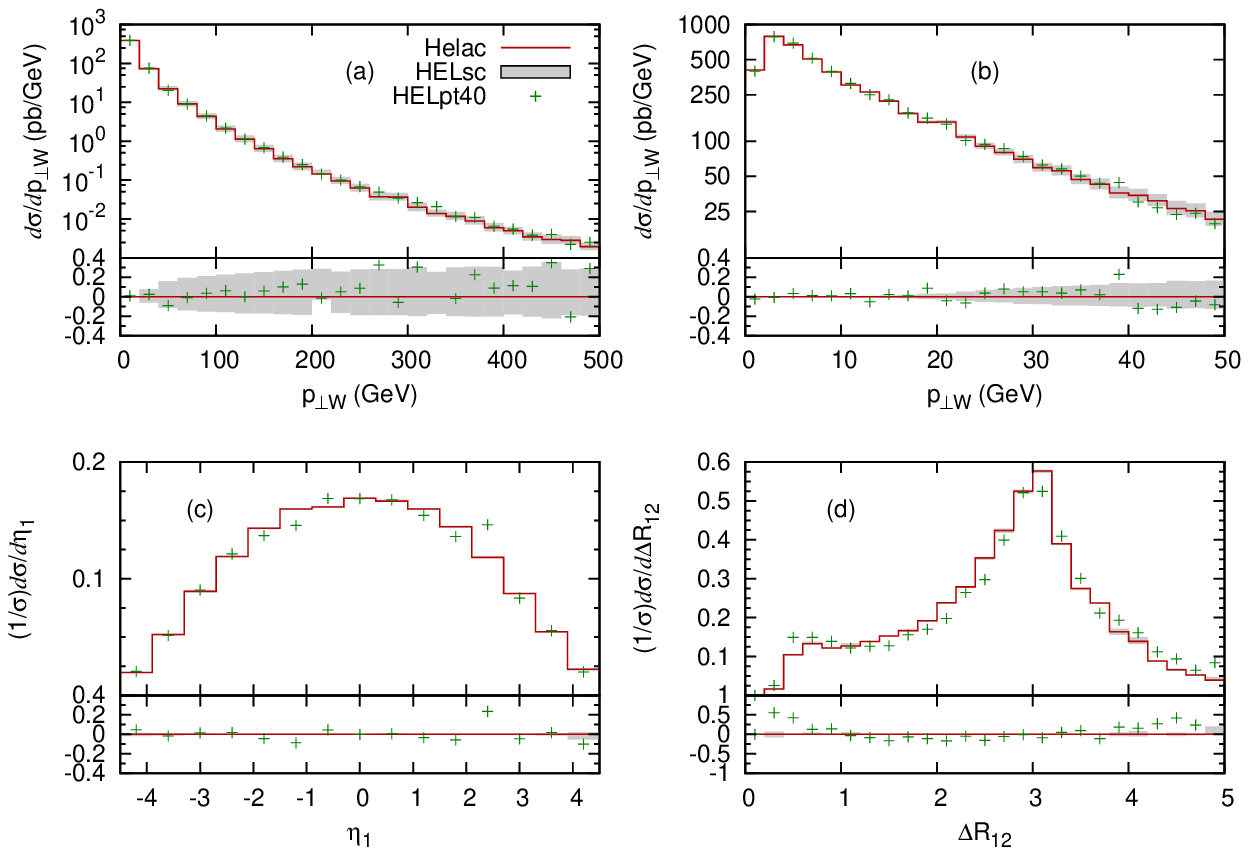}
\includegraphics[width=0.92\textwidth,clip]{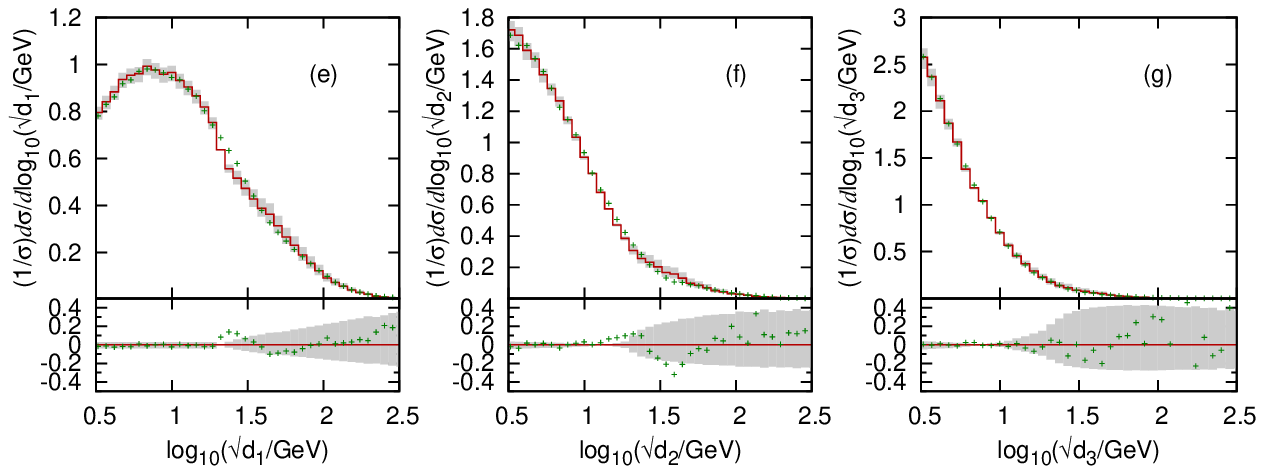}
\end{center}
\vskip -0.4cm
\ccaption{}{\label{fig:hel-ptw-lhc}\protect\helac systematics at the LHC.
  The plots are the same as in \fig{fig:alp-ptw-lhc}. The full line is
  the default settings of \protect\helac, the shaded area is the range
  between HELscL and HELscH, while the points represent HELpt40 as
  defined in section~\ref{sec:genprop}.}
\end{figure}

The Tevatron \helac distributions are shown in \fig{fig:hel-ptw-tev}.
Since \helac results presented in this study are based on the MLM
matching prescription, we expect the \helac systematics to follow at
least qualitatively the \alpgen ones and this is indeed the case. On
the other hand the use by \helac of \pythia, for parton showering as
well as for hadronization, leads to differences compared to the
\alpgen results, where \herwig is used. For the absolute rates,
especially in the multi-jet regime, \helac seems to be closer to
\madevent that also uses \pythia.

For the LHC, the \helac systematics are shown in \fig{fig:hel-ptw-lhc}.
The systematics follows a similar pattern compared to that already
discussed for the Tevatron case, with the expected increase of up to
40\% from scale variations, due to the higher collision energy.

\subsection{\protect\madevent systematics}
\label{sec:mad-syst}

\begin{figure}
\vskip 1.6cm
\begin{center}
\includegraphics[width=0.92\textwidth,clip]{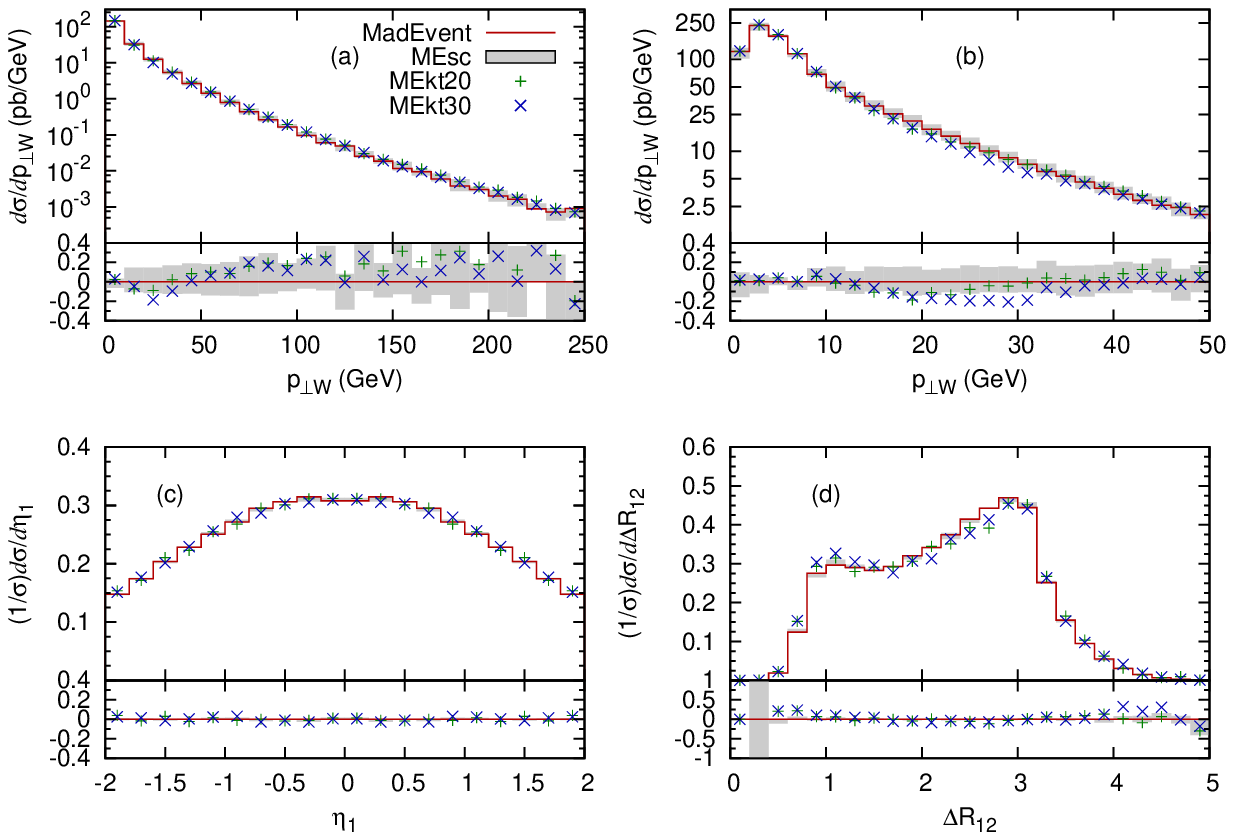}
\includegraphics[width=0.92\textwidth,clip]{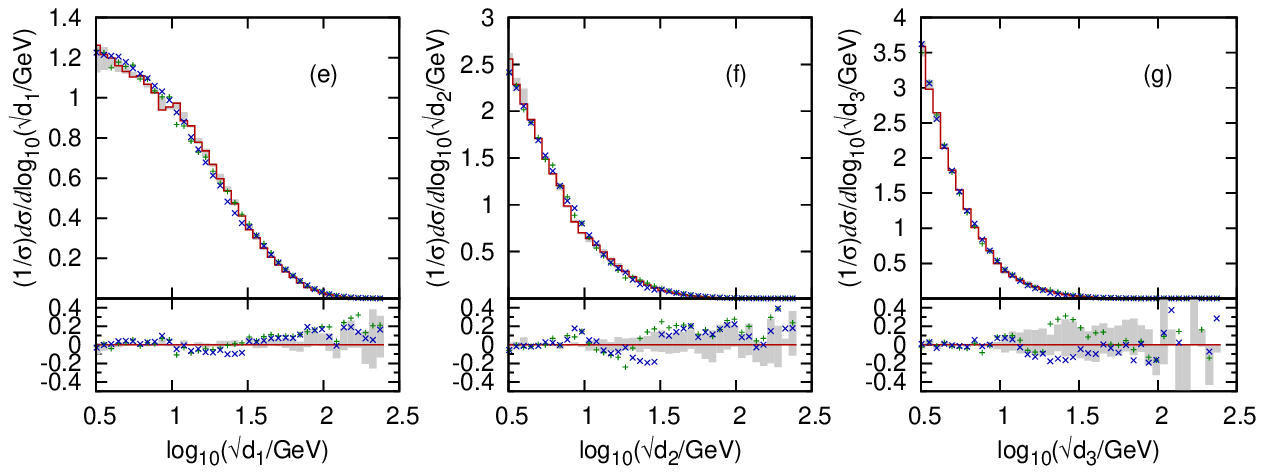}
\end{center}
\vskip -0.4cm
\ccaption{}{\label{fig:mad-ptw-tev}\protect\madevent systematics at the
  Tevatron. The plots are the same as in \fig{fig:alp-ptw-tev}. The
  full line is the default settings of \protect\madevent, the shaded
  area is the range between MEscL and MEscH, while the points
  represent MEkt20 and MEkt30 as defined in section~\ref{sec:genprop}.}
\vskip 1.6cm
\end{figure}

\begin{figure}
\begin{center}
\includegraphics[width=0.92\textwidth,clip]{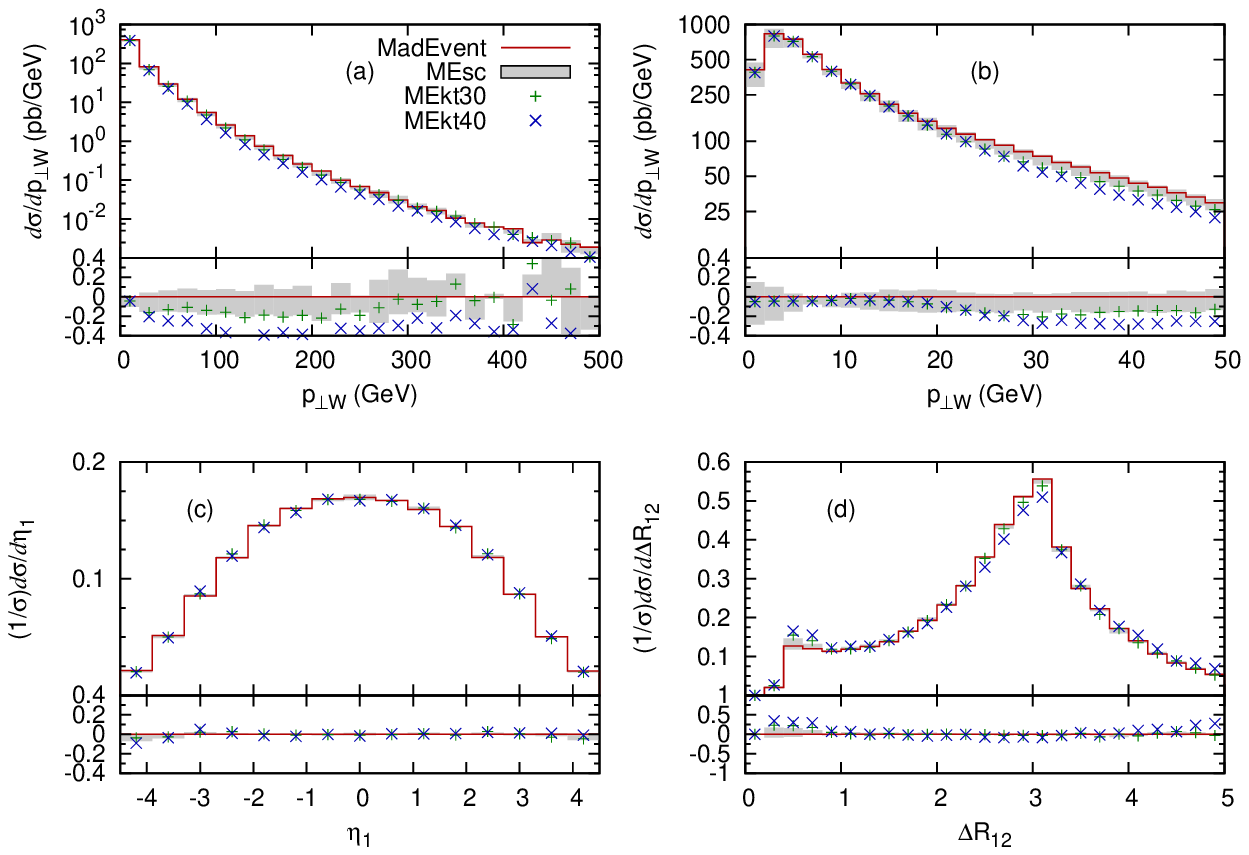}
\includegraphics[width=0.92\textwidth,clip]{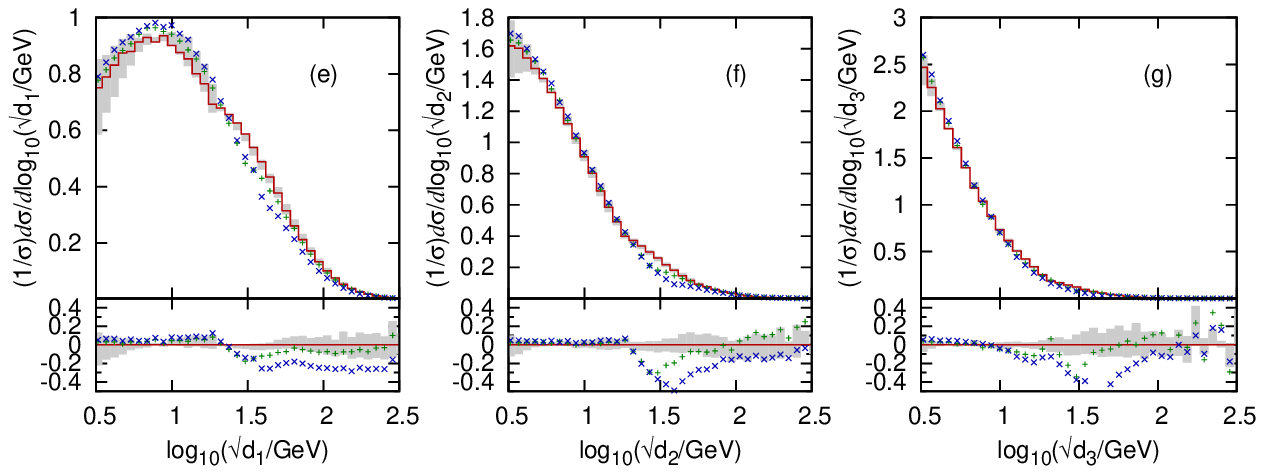}
\end{center}
\vskip -0.4cm
\ccaption{}{\label{fig:mad-ptw-lhc}\protect\madevent systematics at the
  LHC. The plots are the same as in \fig{fig:alp-ptw-lhc}. The full
  line is the default settings of \protect\madevent, the shaded area
  is the range between MEscL and MEscH, while the points represent
  MEkt30 and MEkt40 as defined in section~\ref{sec:genprop}.}
\end{figure}

The \madevent distributions for the Tevatron are shown in
\fig{fig:mad-ptw-tev}.  Also here, the variations are consistent with
the expectations. For the \ppw\ spectrum, the dominant variations are
due to the change of scale for $\alpha_s$, with the lower scale
leading to a harder spectrum. Below the \kperp-cutoff, where the
distribution is determined by the parton shower only, the lower scale
gives the lower differential cross section.

At Tevatron energies, both the \ppw\ spectrum and the $d_i$ spectra are
relatively stable with respect to variations of the matching scale. For
the $d_i$ spectra, the variation in matching scale gives a dip in the
region $10\ \gev < \sqrt{d_i} < k_{\perp 0}$, but is reduced for larger
$d_i$. The rapidity and jet-distance spectra show a remarkable stability
under both renormalization-scale changes and variations in the cutoff
scale.

For the LHC, the systematics of the \madevent implementation are shown
in \fig{fig:mad-ptw-lhc}. The variations in renormalization scale give
a very similar effect as for the Tevatron, with variations up to $\pm
20\%$ on the \pperp\ and $d_i$ spectra. For variations in the matching
scale $k_{\perp 0}$, however, the pattern is slightly different. This
can be most easily understood from looking at the $d_i$ spectra,
since, as in the Sherpa case, the cutoff scale is defined to be just
the $d_i$, so the transition between the parton-shower and
matrix-element regions is very sharp. It is clear from these
distributions that the default parton shower of \pythia does not
reproduce the shape of the matrix elements at LHC energies even for
relatively small \kperp, but falls off more sharply.  There is
therefore a dip in all the distributions around $\log k_{\perp 0}$,
which gets more pronounced for the higher multiplicity distributions,
and hence gives lower overall jet rates. The \ppw\ distributions, as
well as the $d_1$ distributions, are composed of all the different
jet-multiplicity samples, which gives systematically reduced hardness
of the differential cross sections for increased cutoff scales. These
effects are clearly visible also in \sherpa, which uses a \pythia-like
parton shower and \kperp\ as merging scale.

\subsection{\protect\sherpa systematics}
\label{sec:she-syst}

\begin{figure}
\begin{center}
\includegraphics[width=0.92\textwidth,clip]{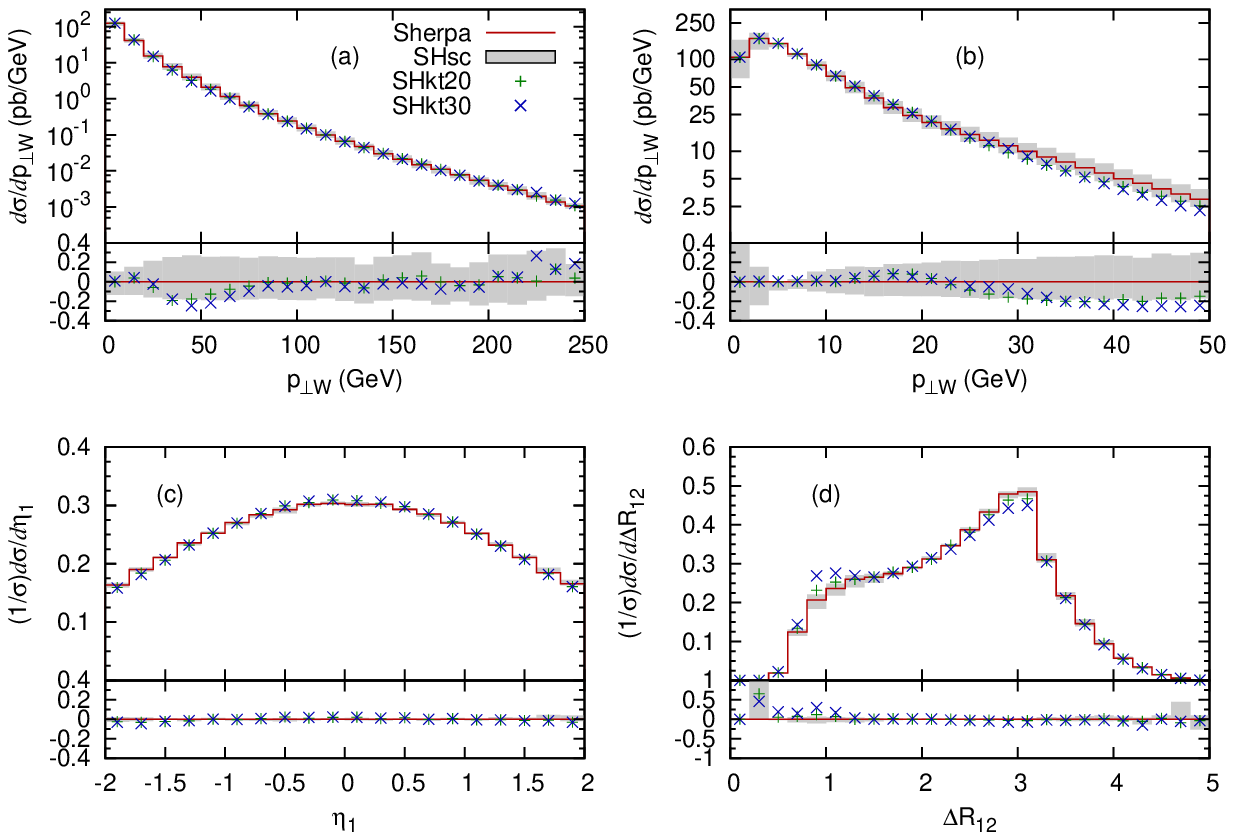}
\includegraphics[width=0.92\textwidth,clip]{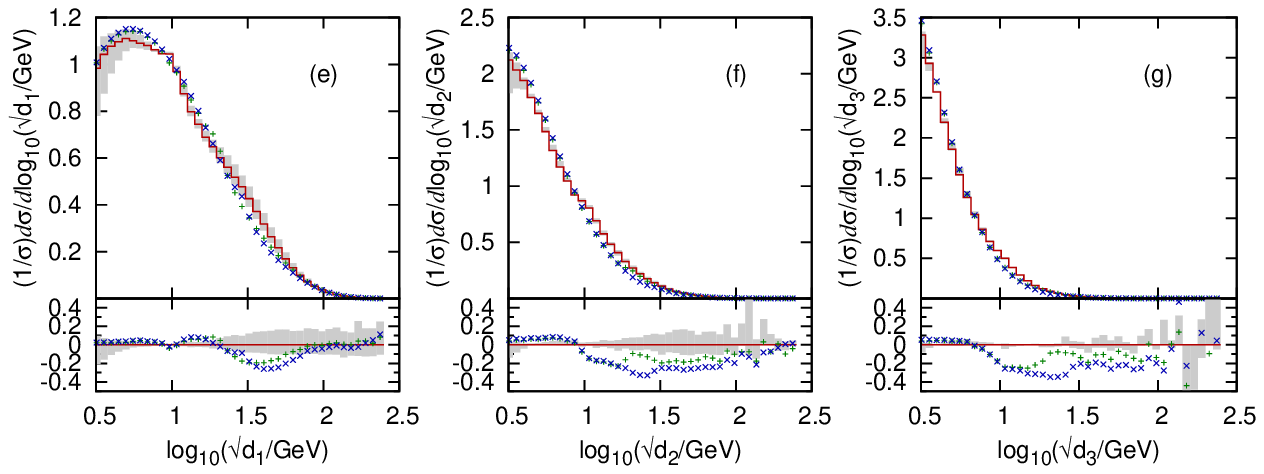}
\end{center}
\vskip -0.4cm
\ccaption{}{\label{fig:she-ptw-tev} \protect\sherpa systematics at the
  Tevatron. The plots are the same as in \fig{fig:alp-ptw-tev}. The
  full line is the default settings of \protect\sherpa, the shaded
  area is the range between SHscL and SHscH, while the points
  represent SHkt20 and SHkt30 as defined in
  section~\ref{sec:genprop}.}
\end{figure}

\begin{figure}
\begin{center}
\includegraphics[width=0.92\textwidth,clip]{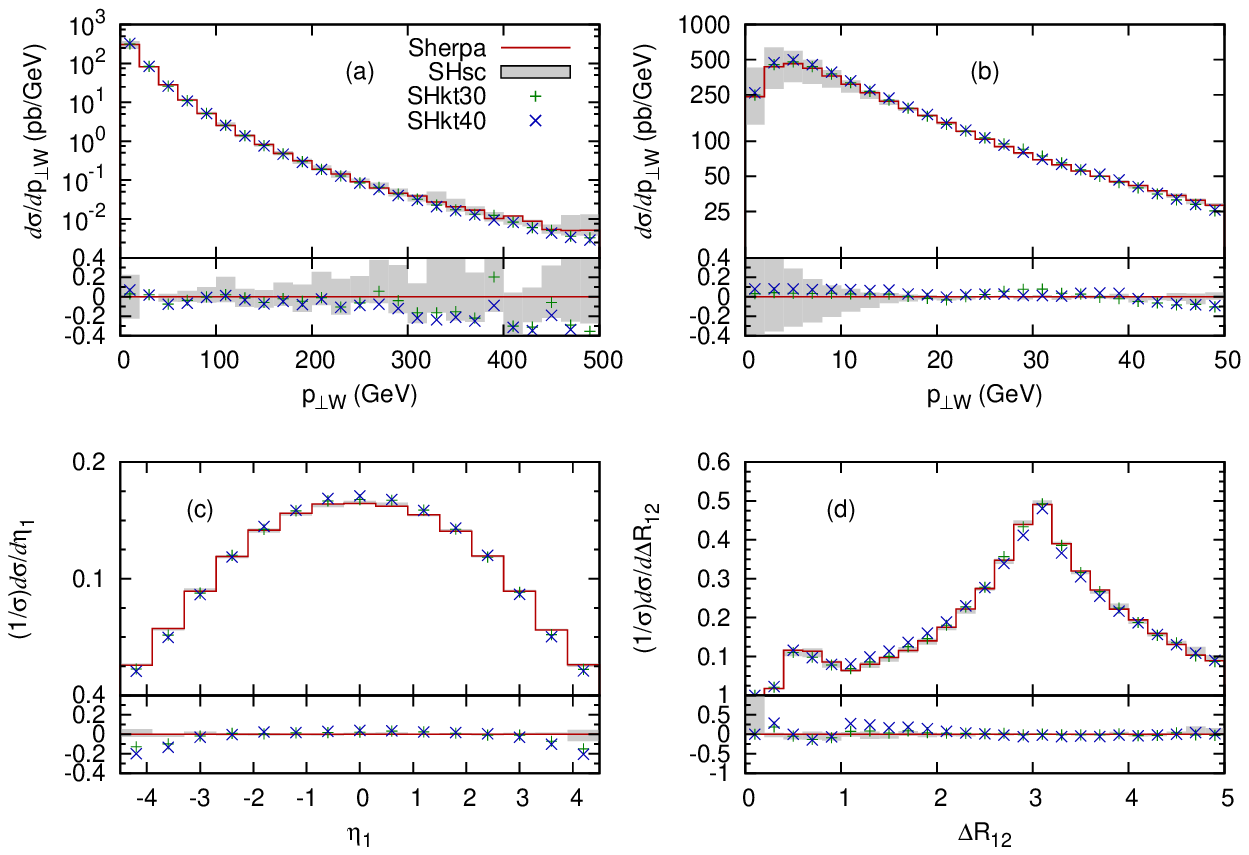}
\includegraphics[width=0.92\textwidth,clip]{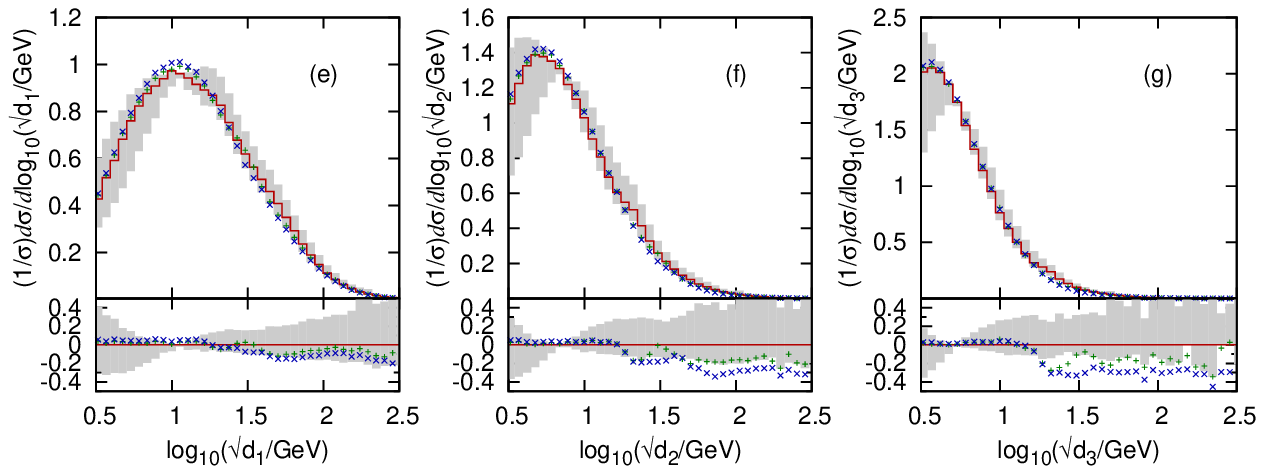}
\end{center}
\vskip -0.4cm
\ccaption{}{\label{fig:she-ptw-lhc}\protect\sherpa systematics at the
  LHC. The plots are the same as in \fig{fig:alp-ptw-lhc}. The full
  line is the default settings of \protect\sherpa, the shaded area is
  the range between SHscL and SHscH, while the points represent
  SHkt30 and SHkt40 as defined in section~\ref{sec:genprop}.}
\end{figure}

The systematics of the CKKW algorithm as implemented in \sherpa is
presented in \fig{fig:she-ptw-tev} for the Tevatron case.  The effect
of varying the scales in the PDF {\it and}\/ strong coupling
evaluations by a factor of $0.5\ (2.0)$ is that for the lower (higher)
scale choice, the $W$-boson's \pperp\ spectrum becomes harder (softer).
For this kind of observables the uncertainties given by scale
variations dominate the ones emerging through variations of the
internal separation cut. This is mainly due to a reduced (enhanced)
suppression of hard-jet radiation through the $\alpha_s$ rejection
weights. The differential jet rates, $d_{1,2,3}$, shown in
\fig{fig:she-ptw-tev}e--g, have a more pronounced sensitivity on the
choice of the merging scale, leading to variations at the $20\%$
level. In the CKKW approach this dependence can be understood since
the \kperp-measure intrinsically serves as the discriminator to
separate the matrix-element and parton-shower regimes. Hence, the
largest deviations from the default typically appear at
$d_i\approx k_{\perp 0}$. However, the results are remarkably smooth,
which leads to the conclusion that the cancellation of the dominant
logarithmic dependence on the merging cut is well achieved. Moreover,
considering the pseudo-rapidity of the leading jet and the cone
separation of the two hardest jets, these distributions show a very
stable behaviour under the studied variations, since they are
indirectly influenced by the cut scale only. The somewhat more
pronounced deviation at low $\Delta R_{12}$ is connected to
phase-space regions of jets becoming close together, which is affected
by the choice of the merging scale and therefore by its variation.
Taken together, \sherpa produces consistent results with relative
differences of the order of or less than $20\%$ at Tevatron energies.

The \sherpa studies of systematics for the LHC are displayed in
\fig{fig:she-ptw-lhc}. Compared to the
Tevatron case, a similar pattern of variations is recognized. The
\pperp\ spectra of the $W^+$ boson show deviations under cut and scale
variations that remain on the same order of magnitude. However, a
noticeable difference is an enhancement of uncertainties in the
predictions for low \pperp. This phase-space region is clearly
dominated by the parton shower evolution, which in the \sherpa
treatment of estimating uncertainties undergoes scale variations in
the same manner as the matrix-element part. Therefore, the estimated
deviations from the default given for low \pperp\ are very reasonable
and reflect intrinsic uncertainties underlying the parton showering.
For the LHC case, the effect is larger, since the evolution is
dictated by steeply rising parton densities at $x$-values that are
lower compared to the Tevatron scenario. The pseudo-rapidity of the
leading jet and the cone separation of the two hardest jets show again
a stable behaviour under the applied variations, the only slight
exception is the regions of high $|\eta_{{\rm jet}1}|$ where, using a
high \kperp-cut, the deviations are at the $20\%$ level. The effect
of varying the scales in the parton distributions and strong couplings
now dominates the uncertainties in the differential jet rates,
$d_{1,2,3}$, which are presented in \fig{fig:she-ptw-lhc}e--g. This
time, owing to the larger phase space, for the low scale choice,
$\mu=\mu_0/2$, the spectra become up to $40\%$ harder, whereas, for
the high scale choice, the spectra are up to $20\%$ softer. The
variation of the internal merging scale does not induce jumps around
the cut region, however it has to be noted that for higher choices,
e.g.\ $k_{\perp 0}=40$ GeV, there is a tendency to predict softer
distributions in the tails compared to the default. To summarize, the
extrapolation from Tevatron to LHC energies does not yield significant
changes in the predictions of uncertainties under merging-cut and
scale variations; for the LHC scenario, they have to be estimated
slightly larger, ranging up to $40\%$. The results are again
consistent and exhibit a well controlled behaviour when applying the
CKKW approach implemented in \sherpa at LHC energies.

Giving a conservative, more reliable estimate, in \sherpa the strategy
of varying the scales in the strong coupling {\it together with}\/ the
scales in the parton densities has been chosen to assess its
systematics. So, to better estimate the impact of the additional scale
variation in the parton density functions, renormalization-scale
variations on its own have been studied as well. Their results show
smaller deviations wrt.\ the default in the observables of this study
with the interpretation of potentially underestimating the systematics
of the merging approach.  Also, then the total cross sections vary
less and become $9095$ pb and $8597$ pb for the low-\ and high-scale
choice, respectively. Note that, owing to the missing simultaneous
factorization-scale variation, their order is now reversed compared to
SHscL and SHscH, whose values are given in table \ref{tab:lhcrates}.
Moreover, by referring to table \ref{tab:lhcratios} the cross-section
ratios for e.g.\ $\sigma^{[\ge 1]}$/$\sigma^{[\mathrm{tot}]}$ now read
$0.26$ and $0.22$ for the low-\ and high-scale choice, respectively.
This once more emphasizes that the approach's uncertainty may be
underestimated when relying on \alps-scale variations only. From table
\ref{tab:lhcrates} it also can be noted that the total inclusive cross
section given by the full high-scale prediction SHscH is -- unlike
\sherpa's default -- close to the \alpgen default. In contrast to the
MLM-based approaches, which prefer the factorization scale in the
matrix-element evaluation set through the transverse mass of the weak
boson, the \sherpa approach makes the choice of employing the merging
scale $k_{\perp 0}$ instead.  This has been motivated in
\cite{Krauss:2002up} and further discussed in \cite{Krauss:2004bs}.
Eventually, it is a good result that compatibility is achieved under
this additional PDF-scale variation for the total inclusive cross
sections, however it also clearly stresses that there is a
non-negligible residual dependence on the choice of the factorization
scale in the merging approaches.

\subsection{Summary of the systematics studies}
\label{sec:sum-syst}

Starting with the \ppw\ spectra, we find a trivial $20-40$\% effect of
the scale changes, with the lower scale leading to a harder spectrum.
In the case of \alpgen and \helac, this only affects the spectrum
above the matching scale, while for \ariadne, \madevent and \sherpa
there is also an effect below, as there the scale change is also
implemented in the parton shower. For all the codes the change in
merging/matching scale gives effects smaller than or of the order of
the change in \alps\ scale. For \ariadne, the change in the soft
suppression parameter (ARs) gives a softer spectrum, which is expected
as it directly reduces the phase space for emitted gluons.

In the $\eta_{{\rm jet}1}$ and $\Delta R_{12}$ distributions the effects of
changing the scale in \alps are negligible. In all cases, changing the
merging/matching scale also has negligible effects on the rapidity
spectrum, while the $\Delta R_{12}$ tends to become more peaked at
small values for larger merging/matching scales, and also slightly
less peaked at $\Delta R_{12}=\pi$. This effect is largest for
\ariadne while almost absent for \helac.

Finally for the $d_i$ distributions we clearly see wiggles of varying
sizes introduced by changing the merging scales. 

\section{Conclusions}
\label{sec:conclusions}

This document summarizes our comparisons of five independent
approaches to the problem of merging matrix elements and parton
showers. The codes under study, \alpgen, \ariadne, \helac, \madevent
and \sherpa, differ in which matrix-element generator is used, which
merging scheme (CKKW or MLM) is used and the details in the
implementation of these schemes, as well as in which parton shower is
used.

We find that, while the three approaches (CKKW, L, and MLM) aim at a
simulation based on the same idea, namely describing jet production
and evolution by matrix elements and the parton shower, respectively,
the corresponding algorithms are quite different. The main
differences can be found in the way in which the combination of
Sudakov reweighting of the matrix elements interacts with the vetoing
of unwanted jet production inside the parton shower.  This makes it
very hard to compare those approaches analytically and to formalise
the respective level of their logarithmic dependence.
In addition, the different showering schemes used by the different
methods blur the picture further. For instance virtuality ordering
with explicit angular vetoes is used in \sherpa as well as in the
\helac and \madevent approach employ \pythia to do
the showering, \pperp ordering is the characteristic feature of
\ariadne, and, through its usage of \herwig it is angular ordering
that enters into the \alpgen merging approach.
However, although the 
formal level of agreement between the codes is not worked out in this 
publication and remains unclear, the results presented in this publication 
show a reasonably good agreement.  This proves that the variety of
methods for merging matrix elements and parton showers can be employed with 
some confidence in vector boson plus jet production.

The comparison also points to differences, in absolute rates as well
as in the shape of individual distributions, which underscore the
existence of an underlying systematic uncertainty. Most of these
differences are at a level that can be expected from merging
tree-level matrix elements with leading-log parton showers, in the
sense that they are smaller than, or of the order of, differences
found by making a standard change of scale in \alps. In most cases the
differences within each code are as large as the differences between
the codes. And as the systematics at the Tevatron is similar to that
at the LHC, it is conceivable that all the codes can be tuned to
Tevatron data to give consistent predictions for the LHC. To carry out
such tunings, we look forward to the publication by CDF and D\O\ of the
measured cross sections for distributions such as those considered in
this paper, fully corrected for all detector effects.

\section*{Acknowledgments}

Work supported in part by the Marie Curie research training networks
``MCnet'' (contract number MRTN-CT-2006-035606) and ``HEPTOOLS''
(MRTN-CT-2006-035505).

C.G.~Papadopoulos and M.~Worek would like to acknowledge support from
the ToK project ALGOTOOLS, ``Algorithms and tools for multi-particle
production and higher order corrections at high energy colliders'',
(contract number MTKD-CT-2004-014319).

M.~Worek and S.~Schumann want to thank DAAD for support through the
Dresden--Crakow exchange programme.

J.~Winter acknowledges financial support by the Marie Curie Fellowship
program for Early Stage Research Training and thanks the
CERN Theory Division for the great hospitality during the funding
period.

J.~Alwall acknowledges financial support by the Swedish Research
Council.


{\raggedright
\bibliography{MCcomp}
}

\end{document}